\documentclass[twocolumn,5p,times]{elsarticle}

\usepackage{amssymb}

\usepackage[modulo]{lineno}

\usepackage[usenames]{color}

\usepackage{nicefrac}
\usepackage{multirow} 
\usepackage{hyperref}
\usepackage{units}
\usepackage{fixltx2e}  
\usepackage{dblfloatfix}

\journal{Astroparticle Physics}

\begin{document}

\begin{frontmatter}

\title{A new method for reconstructing the muon lateral distribution with an array of segmented counters}

\author[ITEDA]{D. Ravignani \corref{cor1}}
\cortext[cor1]{Corresponding author: diego.ravignani@iteda.cnea.gov.ar}
\author[IAFE]{A. D. Supanitsky}
\address[ITEDA]{ITeDA (CNEA, CONICET, UNSAM), Buenos Aires, Argentina.}
\address[IAFE]{Instituto de Astronom\'ia y F\'isica del Espacio (IAFE, CONICET-UBA), Buenos Aires, Argentina.}

\begin{abstract}
Although the origin of ultra high energy cosmic rays is still unknown, significant progress has been achieved in last decades with the construction of large arrays that are currently taking data. One of the most important
pieces of information comes from the chemical composition of primary particles. It is well known that the muon
content of air showers generated by the interaction of cosmic rays with the atmosphere is rather sensitive to primary mass. Therefore, the measurement of the number of muons at ground level is an essential ingredient to infer the cosmic ray mass
composition. In this work we present a new method for reconstructing the muon lateral distribution
function with an array of segmented counters. The energy range from $\unit[4 \times 10^{17}]{eV}$ to
$\unit[2.5 \times 10^{18}]{eV}$ is considered. For a triangular array spaced at \unit[750]{m} we found that \unit[450]{m} is the optimal distance to evaluate the number of muons. The corresponding statistical and systematic uncertainties of the new
and of a previous reconstruction methods are compared. Since the statistical uncertainty of the new reconstruction is less than in the original one, the power to discriminate between heavy and light cosmic ray primaries is enhanced. The detector dynamic range is also extended in the new reconstruction, so events falling closer to a detector can be included in composition studies.   
\end{abstract}


\end{frontmatter}

\section{Introduction}

The physics of cosmic rays is only partially understood. Indeed some basic questions like where they come from, how
they are accelerated, how they propagate in the Galaxy and beyond, and even what particle type they are, have only
tentative answers. Despite the considerable experimental progress accomplished over the last century, most of the key
points are still open. The flux is well measured but it is just one tool to investigate the nature of cosmic rays. Some important clues have to come from the measurements of the primary mass composition and the arrival direction
distribution. In particular composition data are crucial to find the transition between the
galactic and extragalactic components of cosmic rays (see for instance Ref.~\cite{Medina-Tanco:07}) and to
elucidate the origin of the flux suppression at the highest energies \cite{Kampert:13}. 

The measured all particle energy spectrum extends from below $\unit[10^9]{eV}$ to above $\unit[10^{20}]{eV}$, more than eleven orders of magnitude in energy. It can be roughly approximated by a broken power law with four spectral
features including the knee at a few times $\unit[10^{15}]{eV}$, the ankle at $\sim \unit[3\times10^{18}]{eV}$, and the cutoff at $\sim\unit[3\times10^{19}]{eV}$. In addition a second knee was recently reported by the KASCADE-Grande Collaboration at $\sim \unit[10^{17}]{eV}$ \cite{Apel2011}. 
%
In Ref.~\cite{Kampert2012a} the evolution of primary mass composition with energy, measured by several experiments,
has been reviewed. Although systematic uncertainties are large, data from optical detectors show composition changes in regions corresponding to the observed spectral features. Composition seems to become progressively heavier from the first to the second knee. A transition from a heavy to a light composition then appears up to the ankle. Data collected by the Pierre Auger Observatory show a new change that marks the beginning of a transition from light to heavy primaries in the ankle region~\cite{Abraham2010b,icrc2013_lettesier}. This transition is not confirmed by Telescope Array data, which are more compatible with a proton dominated composition \cite{TA:13}. The present Telescope Array statistics is however insufficient to distinguish between the composition profile seen by Auger and a proton dominated case~\cite{Hanlon:13}. In spite of large systematic uncertainties, surface detector data also show a trend from light to heavy primaries above the knee. In addition data from the Auger surface detector 
show a gradual mass number increase above $\unit[10^{19}]{eV}$~\cite{Gamez:11,Pinto:11}, in agreement with its fluorescence detector observations.  

The cosmic ray composition can be inferred indirectly from the atmospheric depth at which the maximum development of an air
shower is reached ($X_{max}$) and from the muon content at ground level (see for instance Ref.~\cite{Supanitsky2009a}).
The $X_{max}$ parameter is observed with fluorescence telescopes and muons are measured with dedicated detectors. It
is very well known that the shower muonic component is also sensitive to high energy hadronic
interactions. Therefore, as shown in Ref.~\cite{Supanitsky2009} simultaneous measurements of $X_{max}$ and muons allow for their testing.   
%
Different muon detector types have been used since the earliest surface arrays~\cite{Kampert2012} and, more recently, by KASCADE \cite{Antoni2003} and Yakutsk \cite{Glushkov2013}. Two different techniques for counting muons have been employed: there are on the one hand {\em analog} detectors, that produce a signal proportional to the number of muons and {\em digital} ones, that are divided into segments and count muons based on the number of segments with a signal. Hybrid detectors combining both techniques have already been used in CASA-MIA~\cite{Sinclair1989} and AGASA~\cite{Hayashida1995}.    

Muon counters are currently being installed in Auger as part of the AMIGA project (Auger Muons and Infill for the Ground Array), an enhancement to extend the observation energy of Auger down to $\unit[10^{17}]{eV}$ and to perform composition 
studies~\cite{Suarez:13}. AMIGA also includes a triangular array of water Cherenkov detectors identical to those used in the surface array but spaced $\unit[750]{m}$ apart, half its distance. Each AMIGA surface detector will be accompanied by 3 nearby plastic scintillator detectors buried at $\sim \unit[2.5]{m}$ to count muons. In the vertical direction, the soil overburden provides a shielding of 20 radiation lengths and entails a muon threshold of $\sim$\unit[1]{GeV} above ground. The AMIGA muon detector will accept events up to $45^\circ$ of zenith angle. Each muon counter has a sensitive surface area of $\sim\unit[10]{m^2}$ and is divided into 64 scintillator strips of equal size. Each strip is \unit[4]{m} long, \unit[4.1]{cm} wide, and \unit[1]{cm} thick. In the AMIGA case the scintillator strip is the generic segment of a segmented detector. The three counters at each array position are equivalent to a single detector divided into $192$ segments that covers $\unit[30]{m^2}$. 
Each strip is fitted lengthwise with a wavelength 
shifting fibre that drives light to a pixel of a 64 multianode PMT. The muon detector has a dead time of \unit[25]{ns} given by the width of the muon pulse. Muons arriving at the same strip closer in time are not resolved. With the same aim as AMIGA, 3 fluorescence telescopes were deployed by the HEAT project~\cite{Mathes2011}. HEAT and the AMIGA $\unit[750]{m}$ surface array are already fully operational. There are also 22 AMIGA muon counters taking data, 16 of $\unit[10]{m^2}$ and 6 of $\unit[5]{m^2}$. The counters are deployed in an hexagon around a central position. Most of them are installed in two positions to compare the detector response in different conditions, as the burial depth for example. The other 5 positions have a single $\unit[10]{m^2}$ counter each. 

Furthermore, Auger is also planning to upgrade its $\unit[1500]{m}$ surface detector to perform detailed
composition analyses at the highest energies~\cite{Kampert:13}. Therefore, in the near future there will be simultaneous
measurements of $X_{max}$ and muons, the two parameters most sensitive to primary mass, starting from the
second knee region up to the highest energies. These detailed measurements, in which muon detectors play a
fundamental role, will allow unprecedented composition analyses that have the potential to make a decisive contribution
in the understanding of the transition between the galactic and extragalactic cosmic ray components, as well as
the nature of the suppression observed at the highest energies.
%
A method for reconstructing the AMIGA muon lateral distribution function (MLDF) was introduced in Ref.~\cite{Supanitsky2008a}. In this work a new reconstruction that improves the AMIGA resolution is presented. The enhancement is made possible by using an exact likelihood function in the MLDF fit. As a result the power to disentangle the primary composition increases. The new method is suitable for any array of segmented counters.

In section \ref{sec:Methods} the original and the new reconstructions are presented. Section~\ref{sec:simulations} describes air shower and 
detector simulations used in this work and section~\ref{sec:saturation} continues with an estimation of the maximum number of muons that a 
segmented detector can measure before saturating. Section~\ref{sec:comparison} contains a comparison of iron reconstructions at 
$E = \unit[10^{18}]{eV}$ and zenith angle $\theta=30^\circ$, which is extended to all simulated energies and zenith angles in 
section~\ref{sec:performance}. In section~\ref{sec:Conf} we review the performance of the new method for different array configurations. 
We conclude in section~\ref{sec:conclusion}.

\section{Methods for reconstructing the muon lateral distribution function}
\label{sec:Methods}

The average number of muons expected in a detector ($\mu$) depends on the muon density
($\rho$), the detector area ($A$) and the zenith angle ($\theta$) according to
\begin{equation}
\mu  =  \rho \, A \, \cos\theta.
\label{eq:mu}
\end{equation} 
The actual number of muons impinging on the detector ($m$) fluctuates event by event following a Poisson distribution with
parameter $\mu$, i.e.,
\begin{equation}
P(m)  =  e^{-\mu} \, \frac{\mu^m}{m!}.
\label{eq:mprob}
\end{equation} 

\noindent One has to make a distinction between expected and actual number of muons, as the aim of a 
counter is estimating a muon density rather than counting the exact number of particles that crossed the
detector. The difference is subtle but important. For example when only one segment has a signal then $m=1$ with
$99.5\%$ of confidence level. The probability of only one segment on when $m=2$, the next best case, is only $0.5\%$.
However in this case, the number of expected muons $\mu$ can only be estimated to fall in the rather broad interval $[0.6,1.7]$ with a
1$\sigma$ confidence level as shown in \ref{sec:appendixA}. The detector resolution is driven by the Poissonian
fluctuations in $m$ rather than by the detector segmentation in this example. In general the resolution is determined by the finite particle number when $m$ is much less than the number of segments. There is a key methodological difference between the original and the new reconstruction. While the first one starts from $m$, the new method is built from $\mu$. Both reconstructions are presented below. A simplified detector model considering the detector size and segmentation is used to compare the reconstructions. This model is a good approximation to the AMIGA detector, which is designed to be close to $100\%$ efficient and to have a low noise level.

\subsection{The original reconstruction method}

The original reconstruction, introduced in~\cite{Supanitsky2008a}, is here briefly reviewed for the sake of completeness. An approximate likelihood function is used by this method. Stations are divided into three classes: saturated, good and silent. A station is silent if it has less than 3 segments on. A segment is on when it has a signal compatible with one or
more muons. The silent limit of 3 was set to reject accidental triggers caused by background muons crossing two strips. A
detector is considered as saturated when 130 or more segments are on in at least one time bin. This limit has to be
set because the adopted likelihood underestimates data errors. At 130 segments the statistical uncertainty is
$\sim 35\%$ larger than the one used in the reconstruction. The underestimation grows with the number of segments on. Finally
 good stations are those that are neither saturated nor silent. The likelihood function is given by,  
\begin{eqnarray}
\mathcal{L}(\vec{p}) \! \! &=& \! \! 
\prod_{i=1}^{N_{sat}} \frac{1}{2} \left[ 1-\textrm{Erf}\left( \frac{n_{i}^{C}-%
\mu(r_i;\vec{p})}{ \sqrt{2 \ \mu(r_i;\vec{p})}} \right) \right] \nonumber \\ 
&& \times \prod_{i=1}^{N_{good}} e^{-\mu(r_i;\vec{p})} \, 
\frac{\mu(r_i;\vec{p})^{n_{i}^{corr}}}{n_{i}^{corr}!} \nonumber \\
&& \times \prod_{i=1}^{N_{sil}} e^{-\mu(r_i;\vec{p})} \, \Bigl(1+%
\mu(r_i;\vec{p})  + \frac{1}{2} \mu(r_i;\vec{p})^2 \Bigr),
%
%
\label{eq:Lik}
\end{eqnarray}
\noindent where the first, second, and third factors correspond to the saturated, good and silent stations, respectively.
$r_i$ is the distance of the $i$-th station to the shower axis. $N_{sat}$, $N_{good}$, and $N_{sil}$ are the numbers of
saturated, good, and silent stations. $n_{i}^{C}$ is the number of segments on and $n_{i}^{corr}$ is the number of muons 
calculated after applying the correction
\begin{equation}
n_{i}^{corr} = \sum_j \frac{\ln \left(1-\frac{k_{j}}{n} \right)}{\ln\left(1-\frac{1}{n}\right)},
\label{eq:nmucorr}
\end{equation} 

\noindent where sum runs over the time bins. $k_{j}$ is the number of segments on in the
$j$-th time bin and $n$ the number of segments, $n=192$ in AMIGA. The MLDF $\mu(r_i;\vec{p})$ depends on the distance between a detector and the shower axis $r_i$ and on free parameters grouped in the vector $\vec{p}$. $\vec{p}$ is obtained by maximising the likelihood function of Eq.~(\ref{eq:Lik}).

\subsection{The new reconstruction method}

\begin{figure} [htp]
\centering
\setlength{\abovecaptionskip}{0pt}
\includegraphics[width=.45\textwidth]{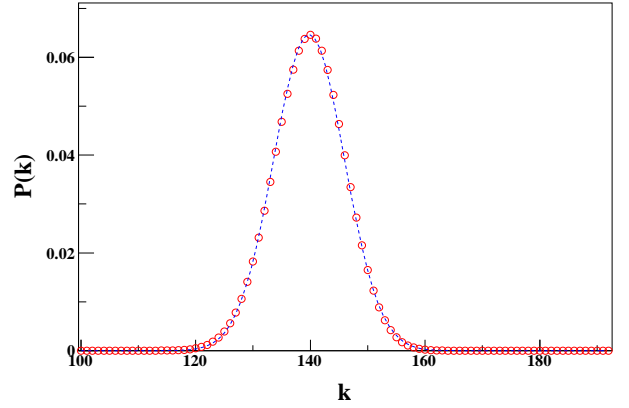}
\caption{Probability of $k$ segments with a signal in a detector divided into 192 segments (open circles). In this example the number of expected muons is 250. The mean value and standard deviation are $139.8$ and $6.2$, respectively. A Gaussian approximation
with the same mean value and standard deviation is shown with a dashed blue line. \label{fig:kprob}}
\end{figure}

The new reconstruction relates the number of segments on to $\mu$. The average number of muons in each segment is $\nicefrac{\mu}{n}$. In  turn the number of muons in a segment follows a Poisson distribution with parameter $\nicefrac{\mu}{n}$. A segment has a signal when one or more muons reaches it and is silent otherwise. Then the probability of a signal, derived from a Poisson distribution, is  $p  =  1 - e^{-\nicefrac{\mu}{n}}$. Calling $k$ the number of segments on, the probability of $k$ as function of $\mu$ follows the binomial distribution 
\begin{eqnarray}
P(k ; \mu) \! \! &=& \! \! \mathcal{L}(\mu ; k) = {n \choose k} \, p^k \,(1-p)^{n-k} \nonumber \\
&=& \! \! {n \choose k} \, e^{-\mu} \, \left( e^{\nicefrac{\mu}{n}} -1 \right)^k. 
\label{eq:kprob}
\end{eqnarray}
Figure \ref{fig:kprob} displays this distribution for AMIGA with $\mu=250$. This example shows that the detector can work well even when there are more muons than segments. The likelihood function from
Eq.~(\ref{eq:kprob}) is used to find the lateral distribution function. For the new reconstruction there is no need to
distinguish between good and saturated detectors since both of them use the same likelihood. As in the
original method, a detector is considered silent when less than 3 segments have signal. The likelihood of silent
detectors is
\begin{eqnarray}
 \mathcal{L}(\mu) \! \! &=& \! \! P(k < 3 ; \mu) \nonumber \\
&=& \! \! e^{-\mu} \, \left(1+n\,\left( e^{\nicefrac{\mu}{n}}-1\right)+\frac{n(n-1)}{2} \left( e^{\nicefrac{\mu}{n}}-1 \right)^2\right).
\label{eq:lsil}
\end{eqnarray} 

The maximum likelihood estimator of the number of expected muons ($\hat{\mu}$) deduced from Eq.~(\ref{eq:kprob}) is   
\begin{equation}
\hat{\mu} = -n \, \ln\left(1-\frac{k}{n}\right). 
\label{eq:muest}
\end{equation}  

\noindent Since $n$ is large $\hat{\mu}$ is close to the estimator of the original reconstruction $n_{i}^{corr}$ in one time bin (see Eq.~\ref{eq:nmucorr}). Note that a single time window is assumed in the new reconstruction, the detector time resolution is not used. 

\section{Numerical simulations}

\label{sec:simulations}

\begin{figure}[h]
\centering
\setlength{\abovecaptionskip}{0pt}
\includegraphics[width=.45\textwidth]{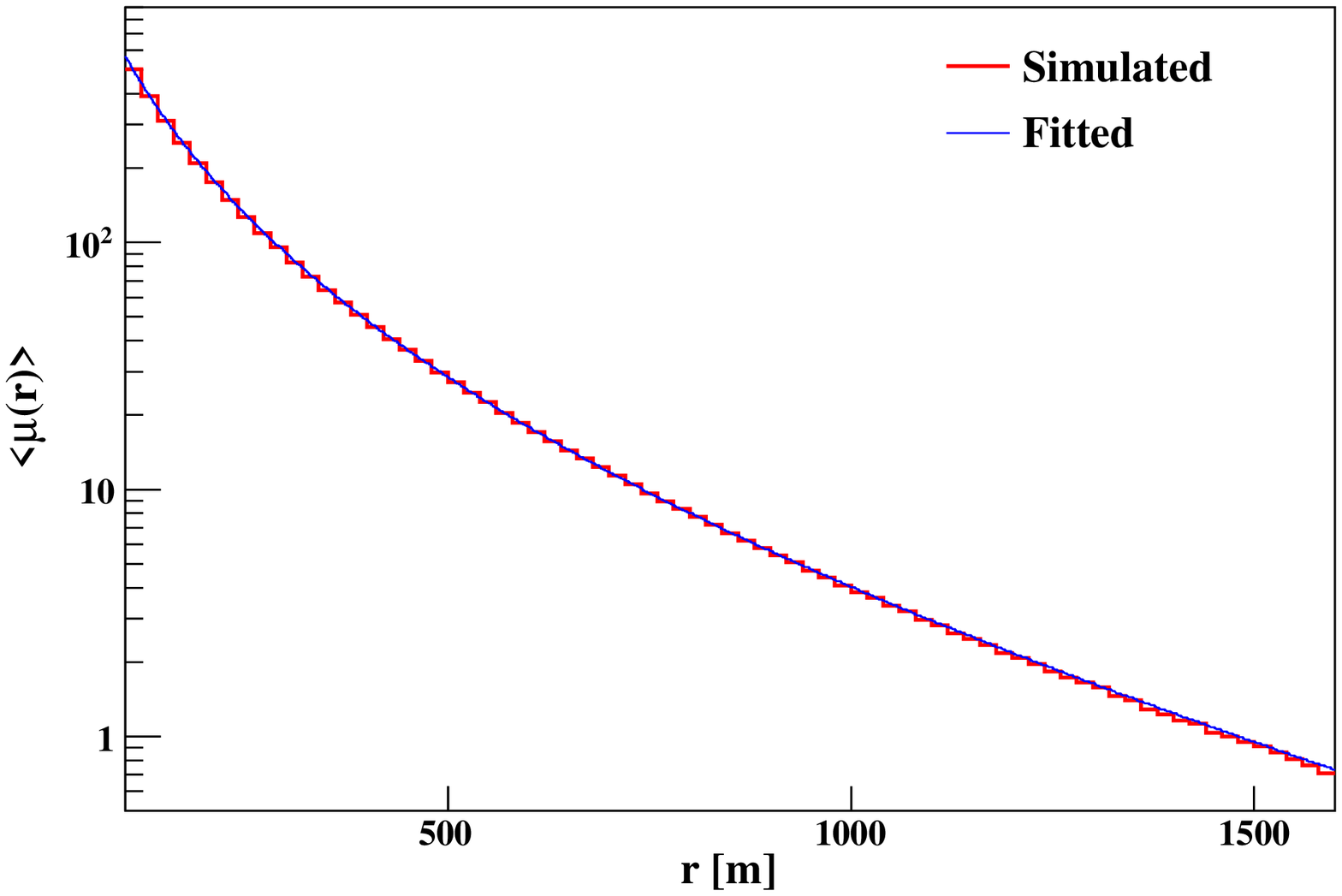}
\includegraphics[width=.45\textwidth]{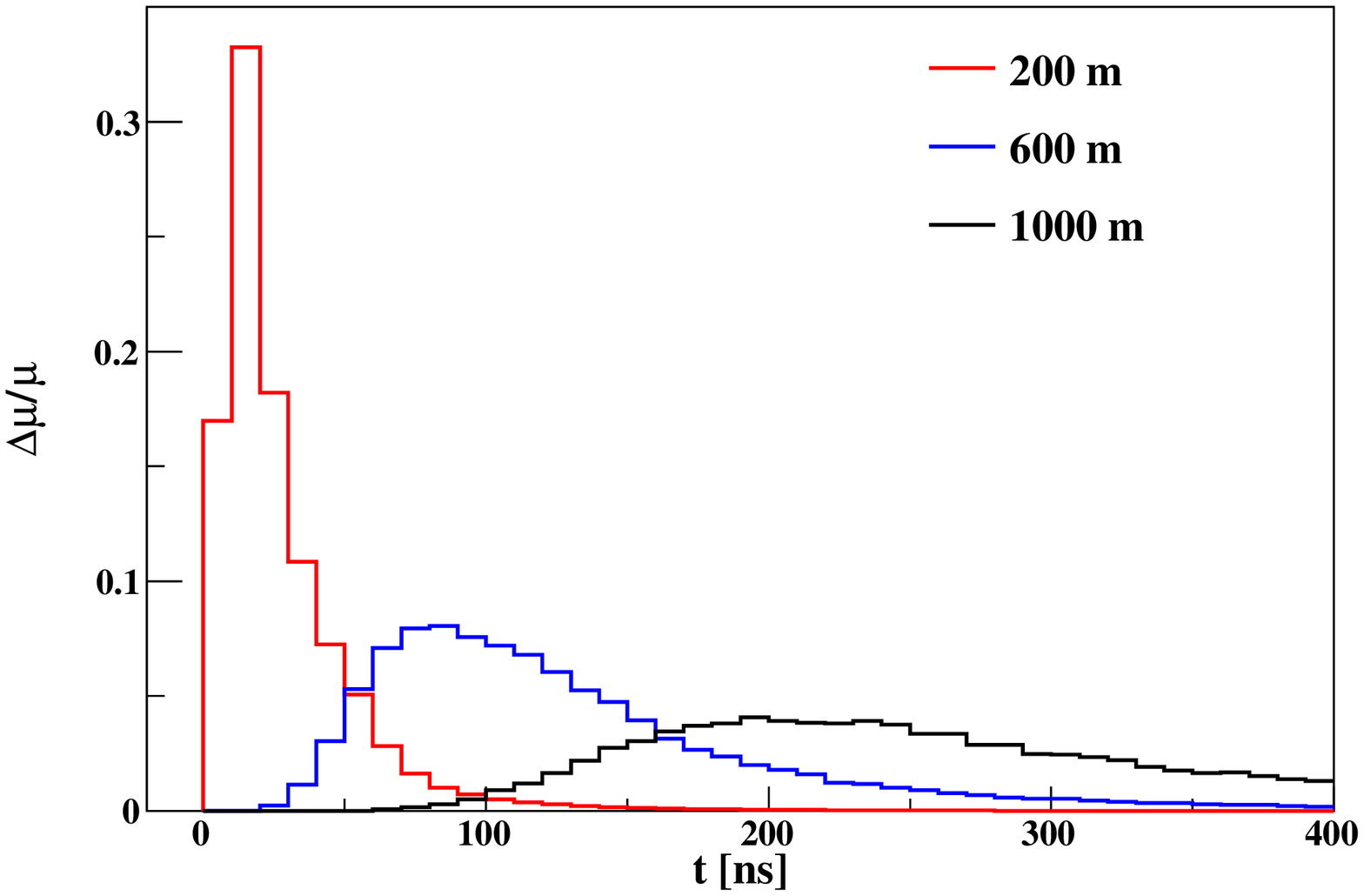}
\caption{Top panel: Average muon lateral distribution function at \unit[2.5]{m} depth obtained from simulations. The
number of muons in a $\unit[30]{m^2}$ counter is represented. A fit with a KASCADE-Grande like function is also shown.
Bottom panel: Muon time distribution at three different shower axis distances. To obtain them 50 iron showers at $E=\unit[10^{18}]{eV}$ and zenith angle $\theta = 30^\circ$ were used. \label{fig:AvSim}}
\end{figure}

A library of atmospheric air showers was generated with AIRES \cite{aires} using QGSJET-II-03 \cite{Ostapchenko2006} as the high energy hadronic 
interaction model. Proton and iron primaries were simulated in the range $\log_{10}(E/\unit{eV}) \in [17.6, 18.4]$ in steps of 
$\Delta \log_{10}(E/\unit{eV}) = 0.2$. The simulations were done for zenith angles $30^\circ$ and $45^\circ$, the median and maximum of the zenith angle distribution respectively. Events at $30^\circ$ represent well the vertical ones. Their main difference is that 
vertical showers have $\approx15\%$ more muons because of the detector projection in the shower plane, which goes as $\sec\theta$. Fifty 
showers were simulated for each primary type, energy, and zenith angle combination. The average MLDF of iron at 
$E=\unit[10^{18}]{eV}$ and $\theta=30^\circ$ is shown in the top panel of Fig.~\ref{fig:AvSim}. A fit with a KASCADE-Grande like MLDF~\cite{KG:10} is also displayed. The number of expected muons in the AMIGA detector ($\mu(r)$) is given by 
\begin{equation}
\label{MLDF}
\mu(r)=A_\mu \left( \frac{r}{r_1} \right)^{-\alpha} \! \left( 1+\frac{r}{r_1} \right)^{-\beta}%
\! \left( 1+\left( \frac{r}{10 \, r_1}\right)^2 \right)^{-\gamma} \! \! \! \!  , 
\end{equation}
where $r$ is the distance to the shower axis, $\alpha = 0.75$, and $r_1 = \unit[320]{m}$. $A_\mu$, $\beta$, and $\gamma$ are free fit parameters. The bottom panel of Fig.~\ref{fig:AvSim} shows the corresponding average muon time distribution at \unit[200]{m}, \unit[600]{m}, and \unit[1000]{m}. As expected for larger $r$ the time distribution becomes wider. The MLDF fit and the muon time distributions obtained for each primary
type, energy and zenith angle are used as input to detector simulations.   

Core positions of simulated events are distributed uniformly. The number of muons
in each station is sampled from a Poisson distribution with a parameter given by the fitted average MLDF. The arrival time of each muon is obtained by sampling the corresponding average time
distribution. The simulation of the pile up of muons in $\unit[25]{ns}$ time bins is also included. A total of $10\,000$ events were simulated for each air shower type.

\section{Saturation}

\label{sec:saturation}

\begin{figure}[htp] 
\centering
\setlength{\abovecaptionskip}{0pt}
\includegraphics[width=.45\textwidth]{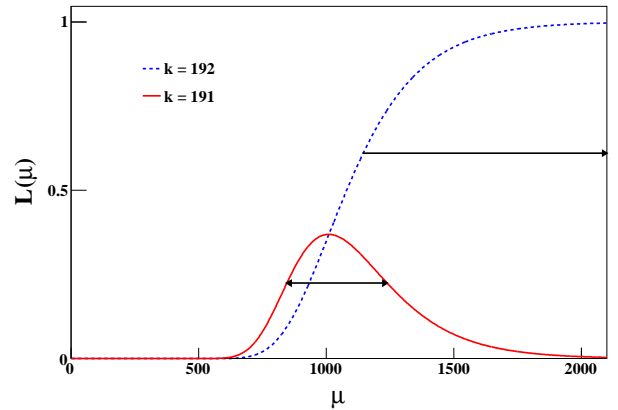}
\caption{Likelihood of the number of expected muons $\mu$ for $k =
191$ and $192$ in a detector consisting of 192 segments. The likelihood reaches a maximum at $\mu=1009$ and infinity
respectively. The intervals where $\ln{L(\mu)}$ falls to 0.5 of its maximum are shown. These intervals are analogous to
the 1$\sigma$ errors of a least squares fit~\cite{Beringer2012}. \label{fig:likesat}}
\end{figure}

We show in this section how segmentation limits the number of muons a detector can count. The likelihood provides an interval
where $\mu$ is allowed to fluctuate in the MLDF fit. If $k$ is less than the number of
segments this interval constrains $\mu$ up and down. The limiting case of all but one segment on is shown in Fig.~\ref{fig:likesat} for the new reconstruction. 
When all segments are on the likelihood becomes a step-like function as shown in Fig.~\ref{fig:likesat}. In this case the effect of the likelihood is to allow $\mu$ to move freely above a lower constraint in the MLDF fit. The lack of an
upper limit to $\mu$ biases the reconstruction. Based on the behaviour of a detector in the fit, a saturation criterion is established. A detector is considered as saturated in the new reconstruction if all segments have a signal and unsaturated otherwise. It is useful to determine the $\mu$ at which the detector saturates ($\mu_{sat}$). Then some saturation criterion in terms of $\mu$ instead of the number of segments on must be established. We choose as $\mu_{sat}$, that which makes the probability of a saturated detector 1\%. From Eq.~(\ref{eq:kprob}), $\mu_{sat}$  is 
\begin{equation}
\mu_{sat}  =  -n \, \ln \left(1-0.01^{\nicefrac{1}{n}} \right).
\label{eq:musat}
\end{equation} 

For the AMIGA detector, consisting of 192 segments, $\mu_{sat} = 719$ muons. Saturated events are excluded from the analysis sample in the new reconstruction due to the aforementioned bias. Note that the rejection of a saturated detector in the fit does not help since it is always better to use it for the lower bound to $\mu$ it provides. 

The fractions of events selected using the saturation criterion are shown in Table~\ref{tab:rsat} for iron at $\theta=30^\circ$. The selection efficiency is higher than $90\%$ at $E=\unit[10^{18.4}]{eV}$, the worst case of all simulated air showers. At same energy, the efficiencies of the other simulated showers are higher than those of iron at $\theta=30^\circ$. A larger muon array than AMIGA, like the \unit[1500]{m} one planned for Auger, is required to collect enough events at energies higher than $\unit[10^{18.4}]{eV}$. The fraction of saturated events for a given shower is, in such more spaced array, lower than in AMIGA.    
 
\begin{table}[t]
\centering
\begin{tabular}{|c|r|r|}
\hline
$\log_{10}(E/\unit{eV})$ & original (\%) & new (\%) \\ \hline \hline
17.6  & 92.6	& 99.5  \\ \hline
17.8  & 92.4	& 99.1	\\ \hline
18.0  & 93.1	& 97.2	\\ \hline
18.2  & 92.3	& 94.8	\\ \hline
18.4  & 89.7	& 90.2	\\ \hline
\end{tabular}
\caption{Selection efficiency of the original and new reconstructions. The case of an iron primary at $\theta=30^\circ$ is shown.}
\label{tab:rsat}
\end{table} 

The likelihood also becomes a step-like function when a station is saturated in the original reconstruction. However the transition occurs 
at $k = 130$ and, applying the 1\% criterion, $\mu_{sat} = 174$~muons. A stricter cut has to be applied in the original than in the new reconstruction. 
Events closer than \unit[100]{m} to the shower axis are excluded in addition to those with all segments on. This distance can be established 
experimentally since the core position uncertainty of the \unit[750]{m} surface array is $\sim$\unit[35]{m} \cite{Medina:06}. The additional 
distance criterion applied in the original reconstruction improves the quality of the analysis sample by reducing the number of events with 
$k \ge 130$, which are biased. 

The original reconstruction efficiency is shown in Table~\ref{tab:rsat}. Differences between the new and the 
original reconstructions are greater at low energies, where the distance cut is more powerful than the saturation one. Events with $k \sim n$ are reconstructed with less quality than the rest. However we selected them in the new reconstructions because differences in resolution are low between these events and the rest. Since most of these events are excluded by the distance cut in the original reconstruction, a better quality event sample is used in the original reconstruction than in the new one.
 
\begin{figure}[htp]
\centering
\setlength{\abovecaptionskip}{0pt}
\includegraphics[width=.45\textwidth]{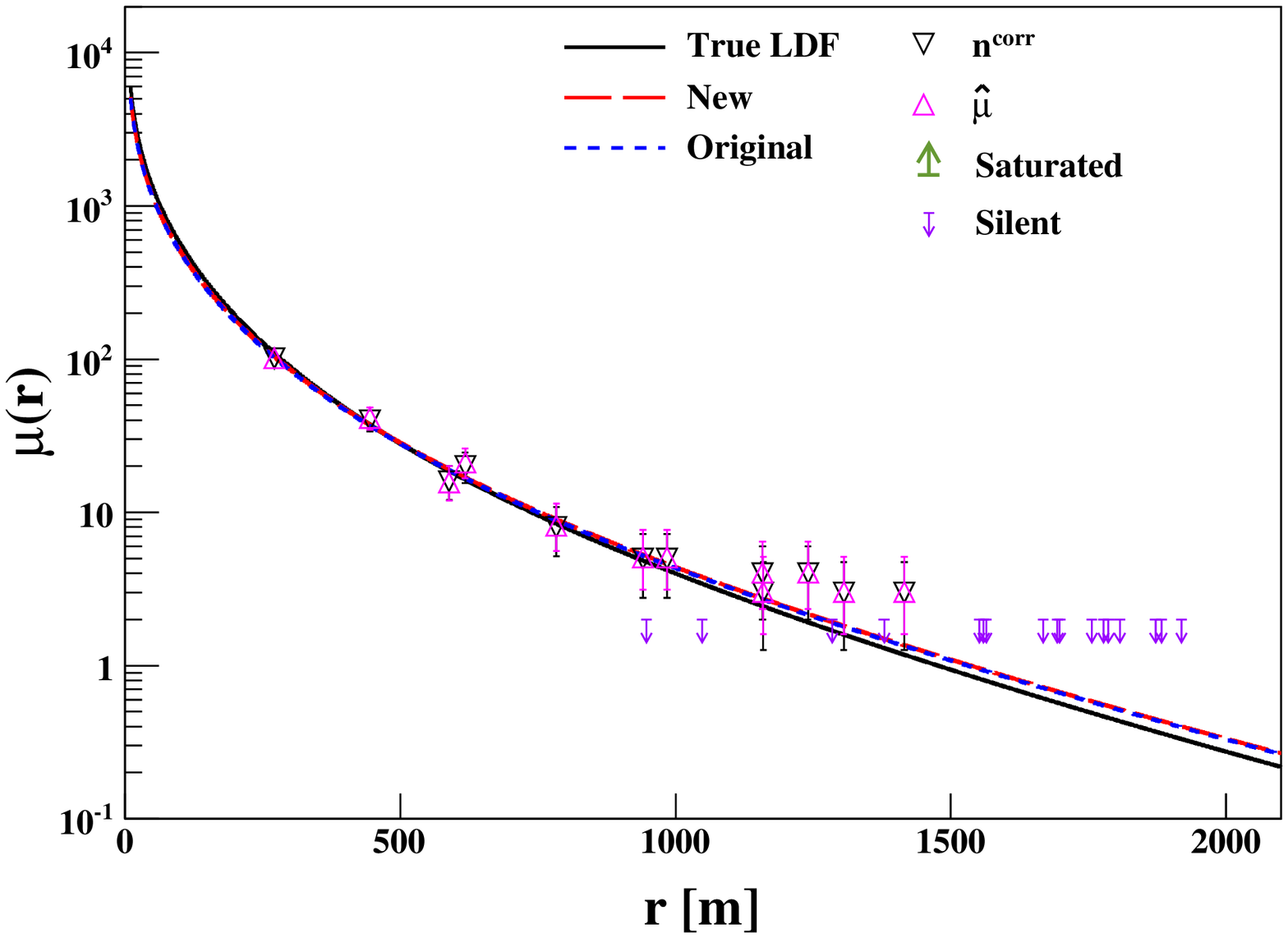}
\includegraphics[width=.45\textwidth]{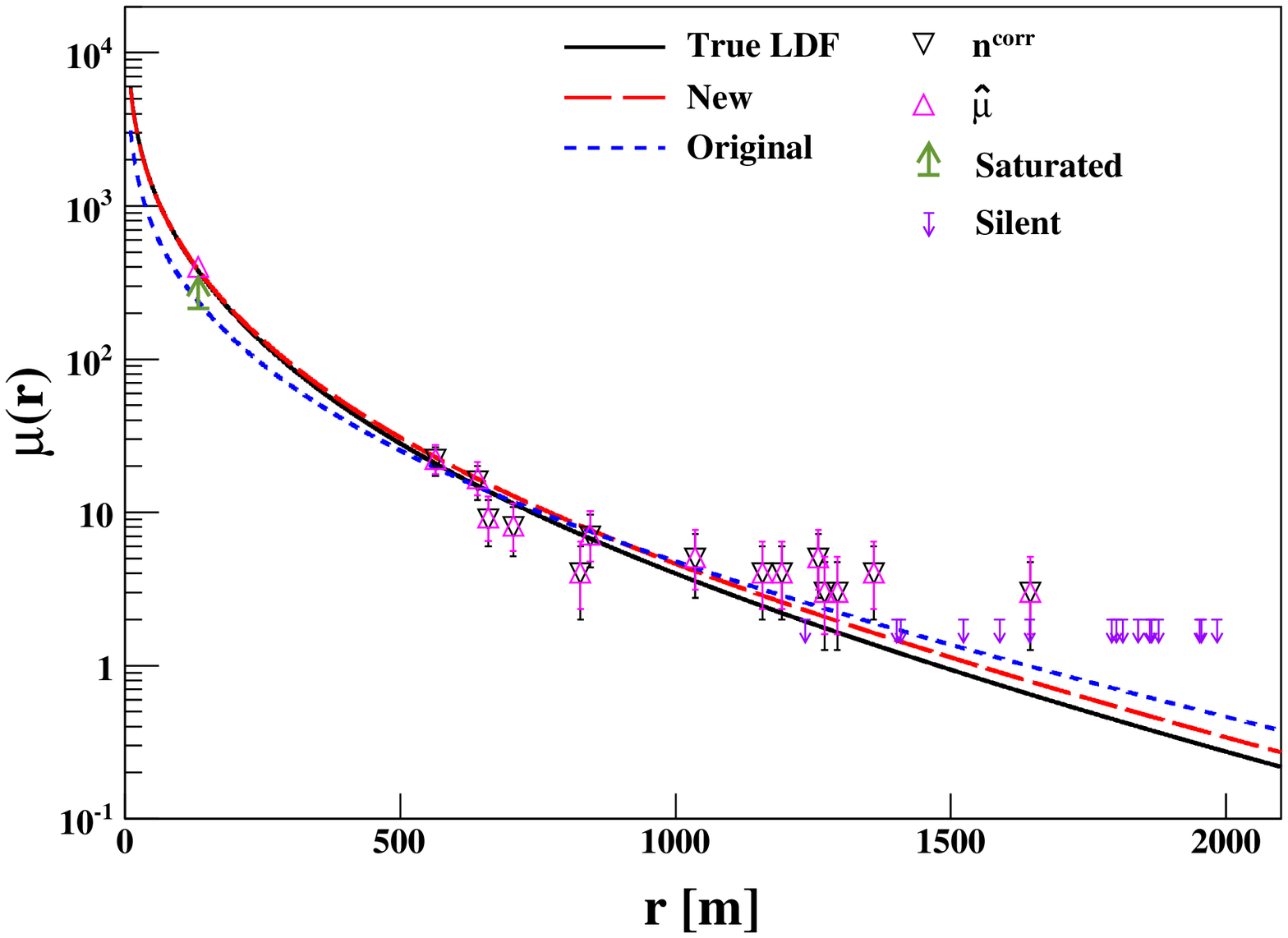}
\caption{Fits of two simulated events using the new and the original reconstructions. The black solid line corresponds to the true MLDF, the dashed red line to the new method, and the dotted blue line to the original method. The event in the top panel has no saturated stations and the one in the bottom panel has a saturated station in the original reconstruction. The down and
up triangles correspond to muons estimated in the original and new reconstructions respectively. Down arrows correspond
to silent stations and the upward arrow to a saturated one. The two events shown correspond to iron showers at
$E=\unit[10^{18}]{eV}$ and zenith angle $\theta = 30^\circ$.
 \label{fig:Fits}}
\end{figure}

\section{Comparison of the reconstruction methods}

\label{sec:comparison}

\begin{figure*} [tp]
\centering
\setlength{\abovecaptionskip}{0pt}
\includegraphics[width=.45\textwidth]{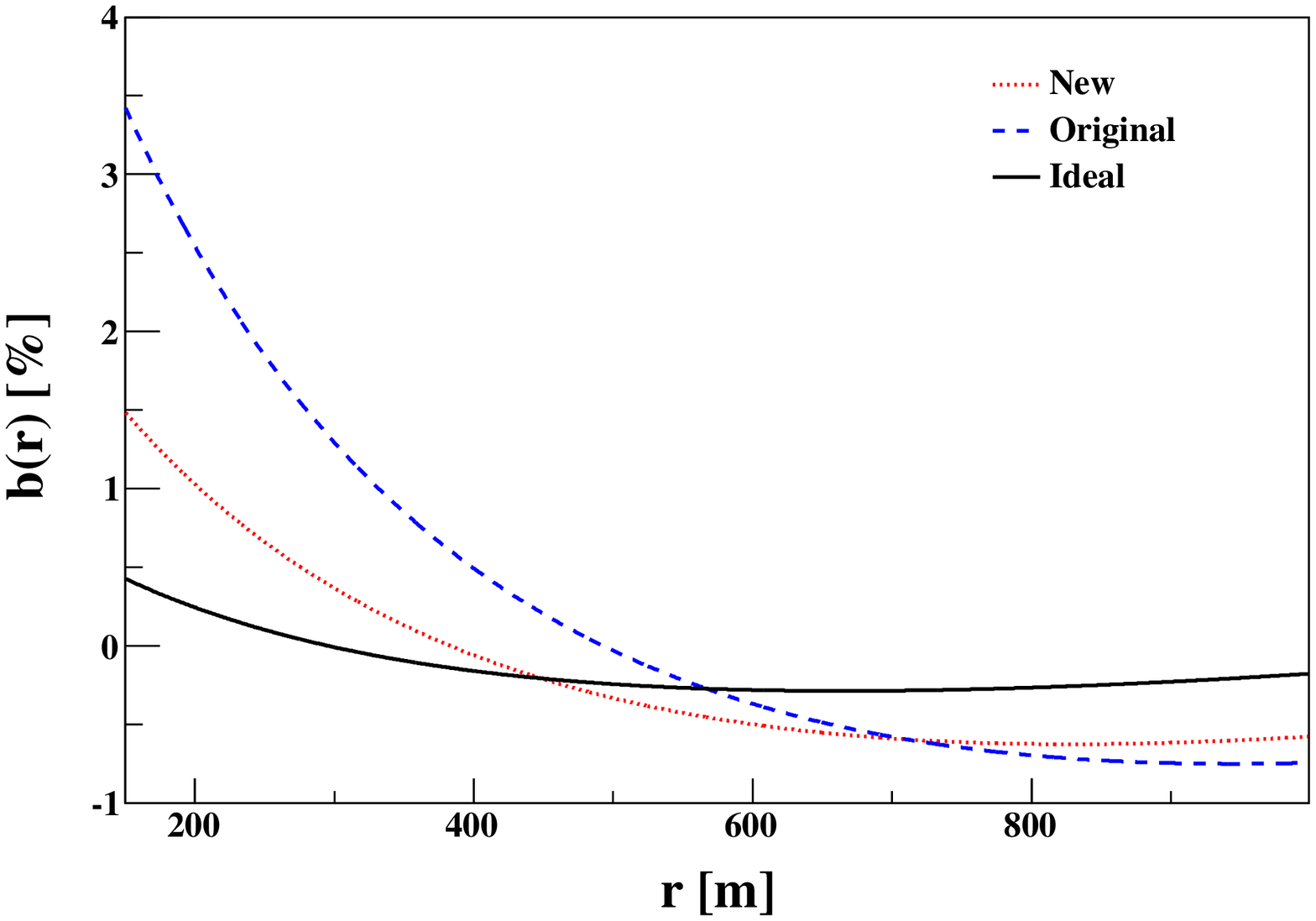}
\includegraphics[width=.45\textwidth]{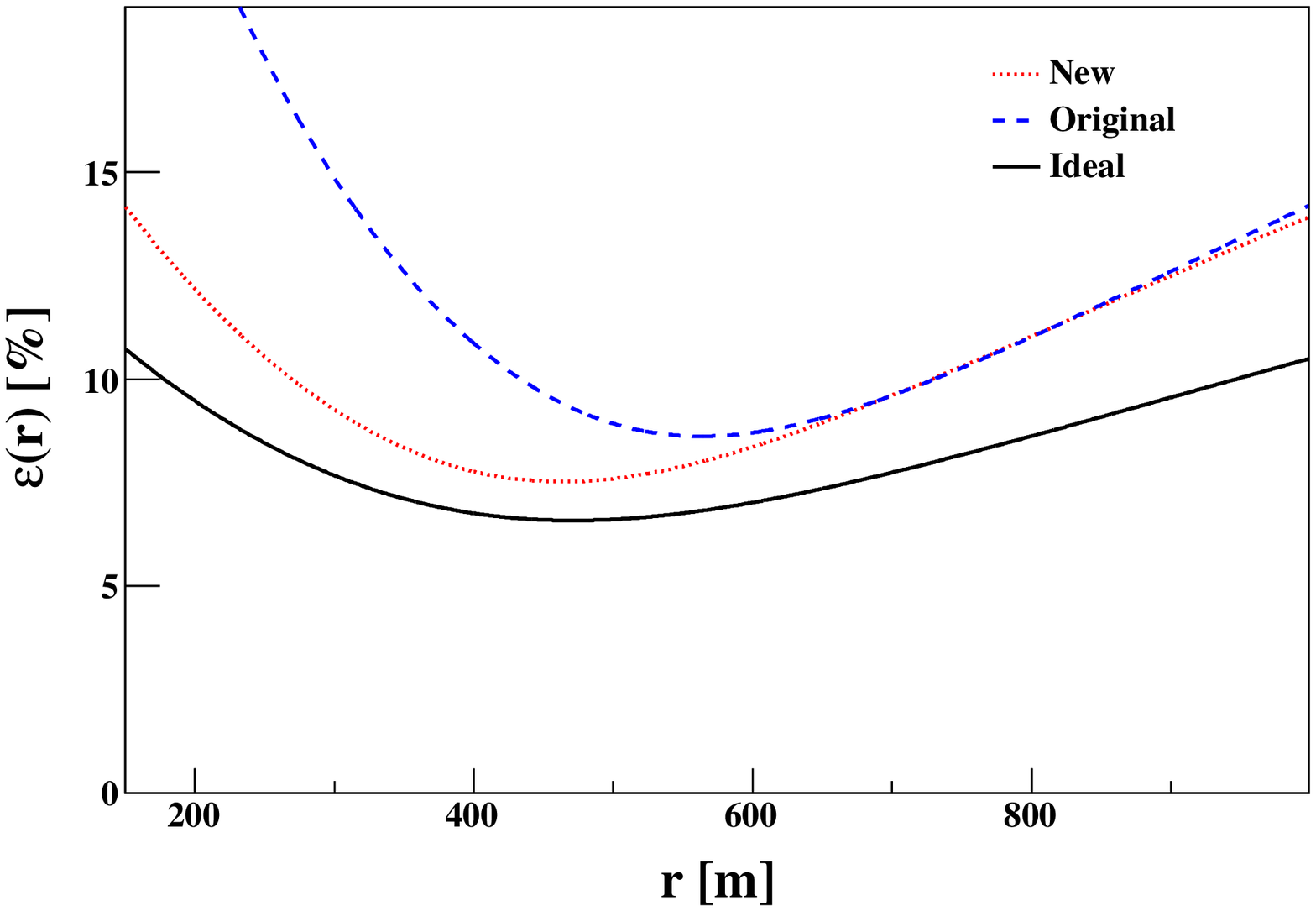}
\caption{Relative bias (left) and standard deviation (right) as a function of the distance to the shower axis.   
An iron primary at $E=\unit[10^{18}]{eV}$ and zenith angle $\theta = 30^{\circ}$ is shown. Events were selected 
according to the quality cuts described in section~\ref{sec:saturation}.  
\label{fig:EpsBvsR_A}}
\end{figure*}

The performances of the original and new methods are compared using simulated events. The reconstruction of iron primaries at $E=\unit[10^{18}]{eV}$ and $\theta =30^\circ$ are presented in this section. An ideal detector of $\unit[30]{m^2}$, the AMIGA size, that counts the number of crossing muons is also considered. The only source of uncertainty arises in this case from the finite number of particles in the detector. The ideal counter sets a lower bound to the resolution achievable with a muon detector. The MLDF is reconstructed for the ideal detector using the likelihood of Eq.~(\ref{eq:mprob}) for any number of muons, no separation in silent / non-silent stations is made. 

For all reconstructions data are fitted with the KASCADE-Grande like function of Eq.~(\ref{MLDF}). In the fit parameters $A_\mu$ and $\beta$ are left free and $\alpha$ and $r_1$ are fixed to the average MLDF values. The parameter $\gamma$, free in the average MLDF fit, is fixed in the reconstruction for three reasons. First, there are not enough triggered detectors to fit it at low energy. Second, $\gamma$ is almost constant with energy, zenith angle, and primary type \cite{Supanitsky2008a}. And last,  $\gamma$ just provides a correction for large core distances, regions where detectors have few muons or are silent. In the showers simulated in this work $\gamma$ varies between $2.9$ and $3$. The average of these two values, $\gamma=2.95$, is used in all reconstructions. Figure \ref{fig:Fits} shows the fits of two iron events. An event with no saturated stations is shown in the top panel. Another one with a saturated station in the original reconstruction, but not in the new one, is displayed in the bottom 
panel. 
Although both methods give very similar fits in the first example, the new method fit is closer to the true MLDF in the second case.

The bias and standard deviation of the MLDFs reconstructed with both methods are compared.
The bias is the systematic uncertainty in the estimation of $\mu(r)$ and the
standard deviation is related, via a confidence interval, to the corresponding statistical uncertainty. The bias of the
fitted $\hat{\mu}(r)$ relative to the number of muons is estimated with
\begin{equation}
b(r)  = \frac{\langle\hat{\mu}(r)\rangle}{\mu(r)} - 1,
\label{eq:bias}
\end{equation}
\noindent where $\langle\hat{\mu}(r)\rangle$ is an average calculated over all reconstructions of the same air shower with
\begin{equation}
\langle \hat{\mu}(r) \rangle  = \frac{1}{N} \sum_{i=1}^{N} \hat{\mu}_i(r),
\label{eq:muhatav}
\end{equation}
\noindent where $N=10\,000$ is the number of simulated events and $\hat{\mu}_i(r)$ is the MLDF reconstructed for the $i$-th event. The standard deviation of $\hat{\mu}(r)$ relative to $\langle\hat{\mu}(r)\rangle$ is estimated with
\begin{equation}
\varepsilon(r)  = \sqrt{ \frac{\sum_{i=1}^{N} \left(\frac{\hat{\mu}_i(r)}{\langle\hat{\mu}(r)\rangle} - 1\right)^2}{N-1} }.
\label{eq:eps}
\end{equation}
\noindent Fig.~\ref{fig:EpsBvsR_A} shows $b(r)$ and $\varepsilon(r)$ of an
iron primary at $E=\unit[10^{18}]{eV}$ and $\theta=30^\circ$. Both reconstruction methods and the ideal
counter are included in the comparison. $\varepsilon(r)$ has a minimum at different distances
$r_{min}$ for each reconstruction. The $r_{min}$ of the original reconstruction is larger than in the new one because, in the first case, saturated stations only contribute with a lower limit to $\mu$. Fluctuations at small distances are therefore increased.  $\varepsilon(r_{min})$ and the bias modulus are smaller in the new
reconstruction than in the original one. The $r_{min}$ and the $\varepsilon(r_{min})$ of the new
reconstruction are close to the ideal detector.

\begin{figure*} [tp]
\centering
\setlength{\abovecaptionskip}{0pt}
\includegraphics[width=.45\textwidth]{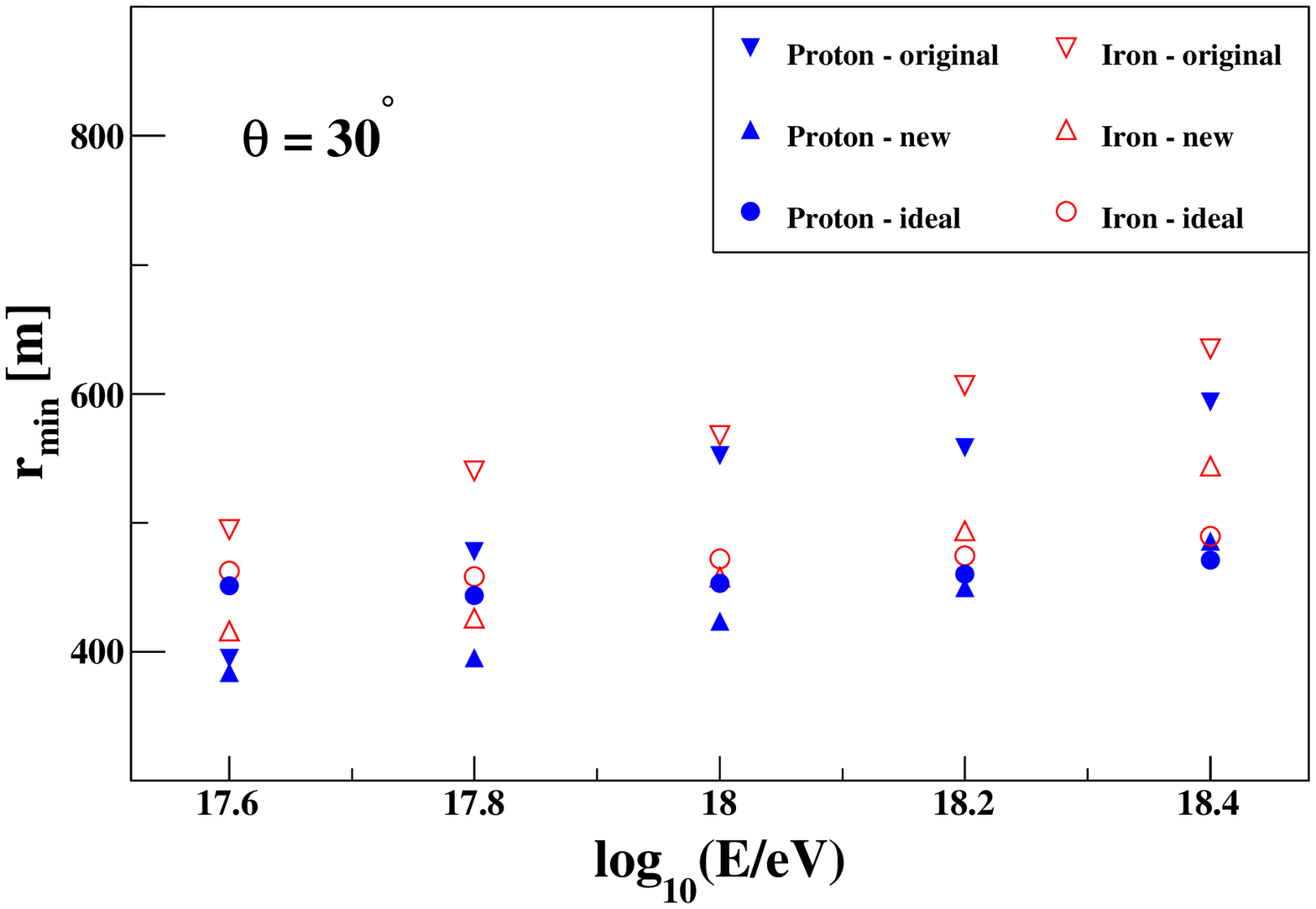}
\includegraphics[width=.45\textwidth]{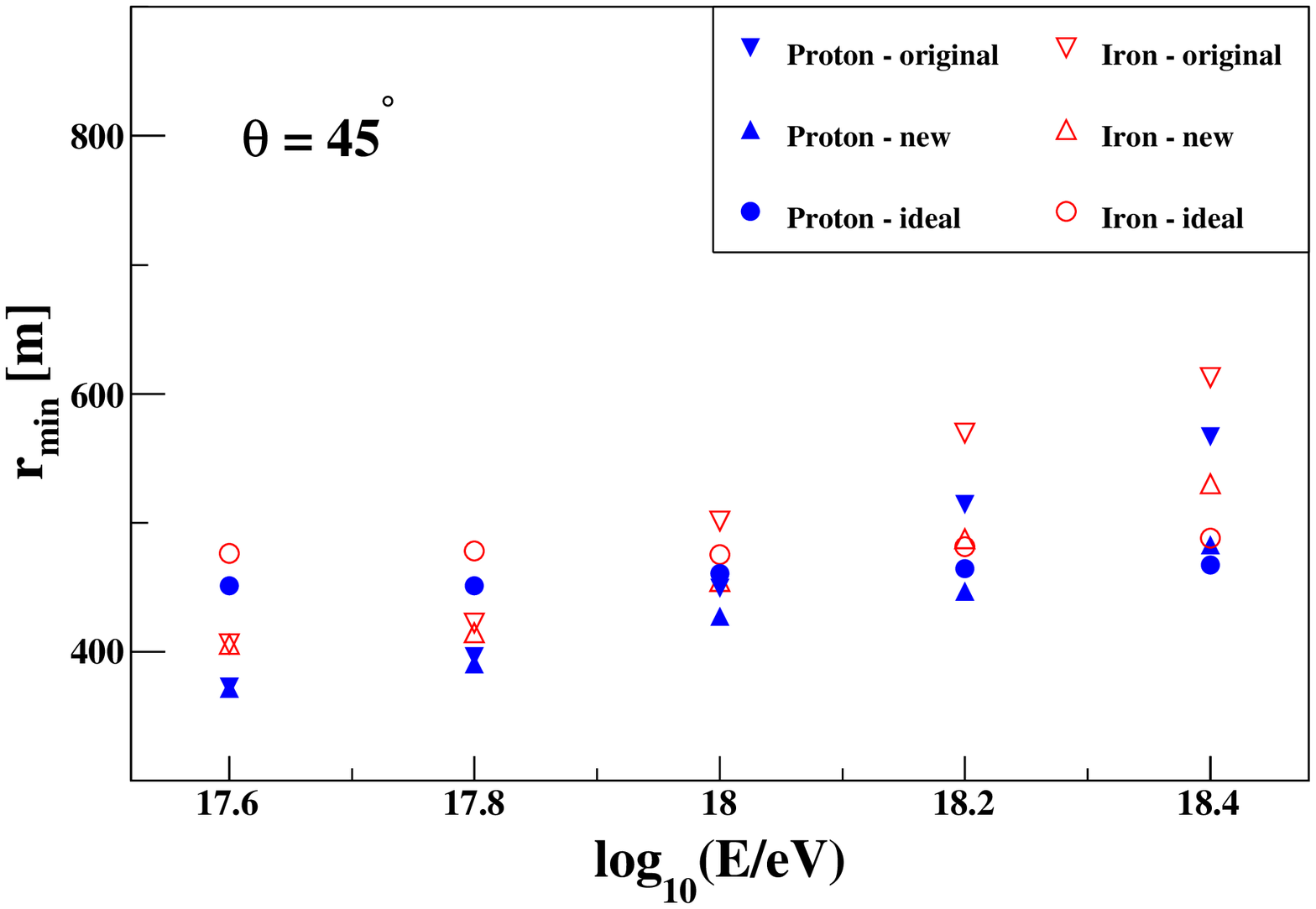}
\caption{Distance at which the statistical uncertainty of the number of muons is minimised ($r_{min}$). Air showers reconstructed with the original reconstruction, the new one, and an ideal muon counter are shown. Zenith angles $30^\circ$ and $45^\circ$ are displayed in the left and right panels respectively. \label{fig:rmin}}
\end{figure*} 

\begin{figure*} [tp]
\centering
\setlength{\abovecaptionskip}{0pt}
\includegraphics[width=.45\textwidth]{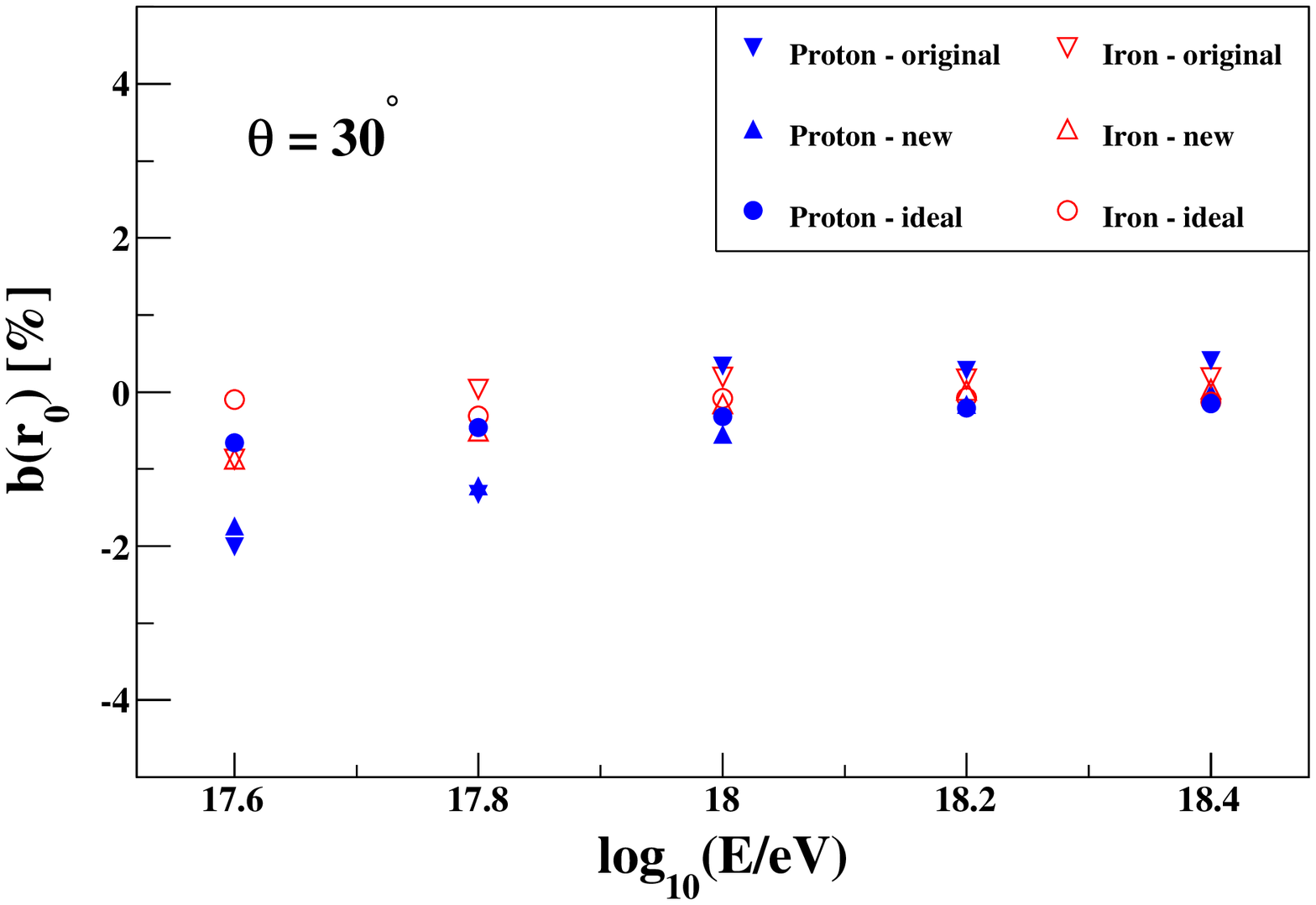}
\includegraphics[width=.45\textwidth]{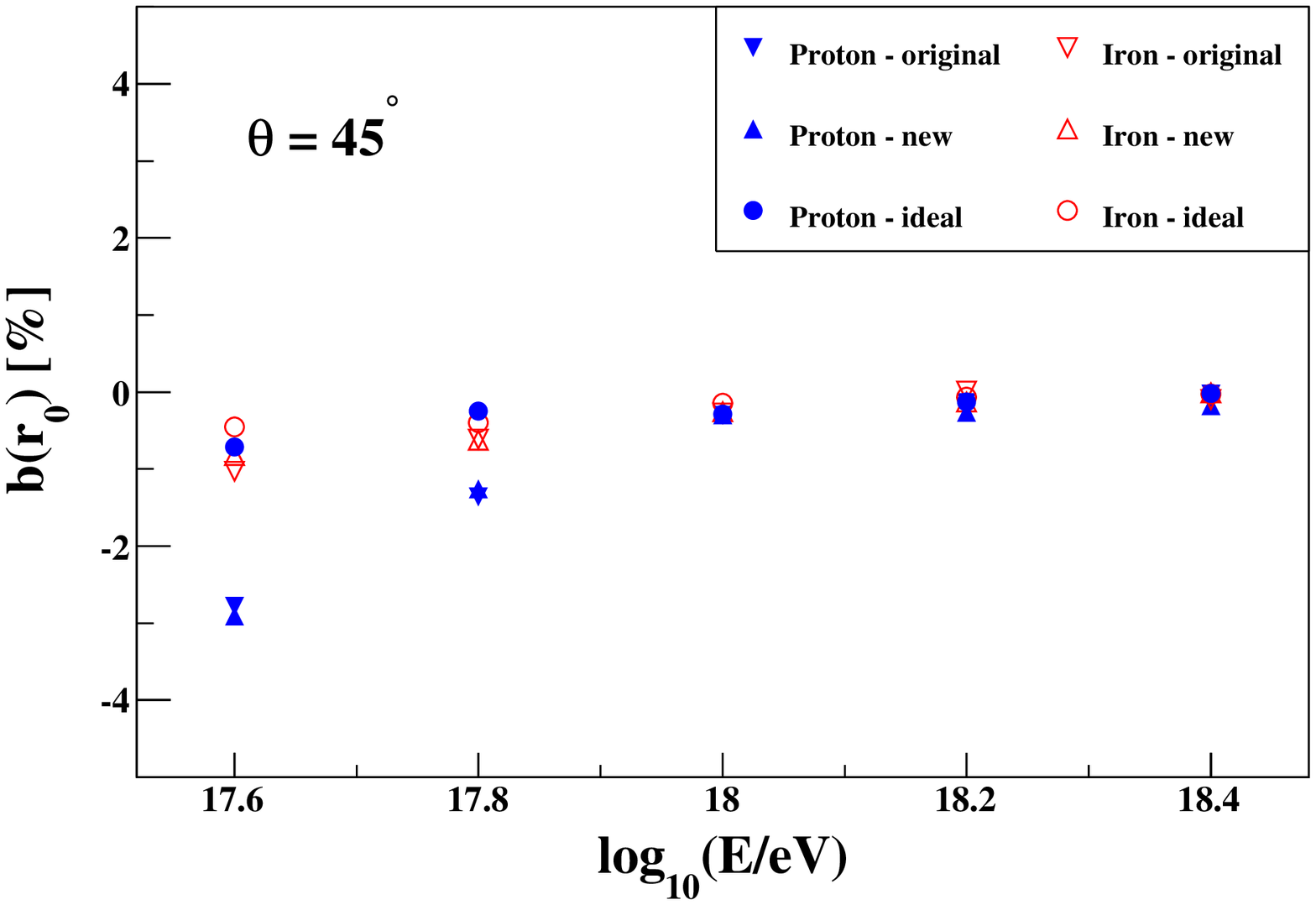}
\caption{Relative bias of the number of muons at $r_0 = \unit[450]{m}$ from the shower axis. Air showers reconstructed with the original reconstruction, the new one, and an ideal muon counter are shown. Zenith angles $30^\circ$ and $45^\circ$ are displayed in the left and right panels respectively.
\label{fig:bias}}
\end{figure*} 

\begin{figure*}[hbp]
\centering
\setlength{\abovecaptionskip}{0pt}
\includegraphics[width=.45\textwidth]{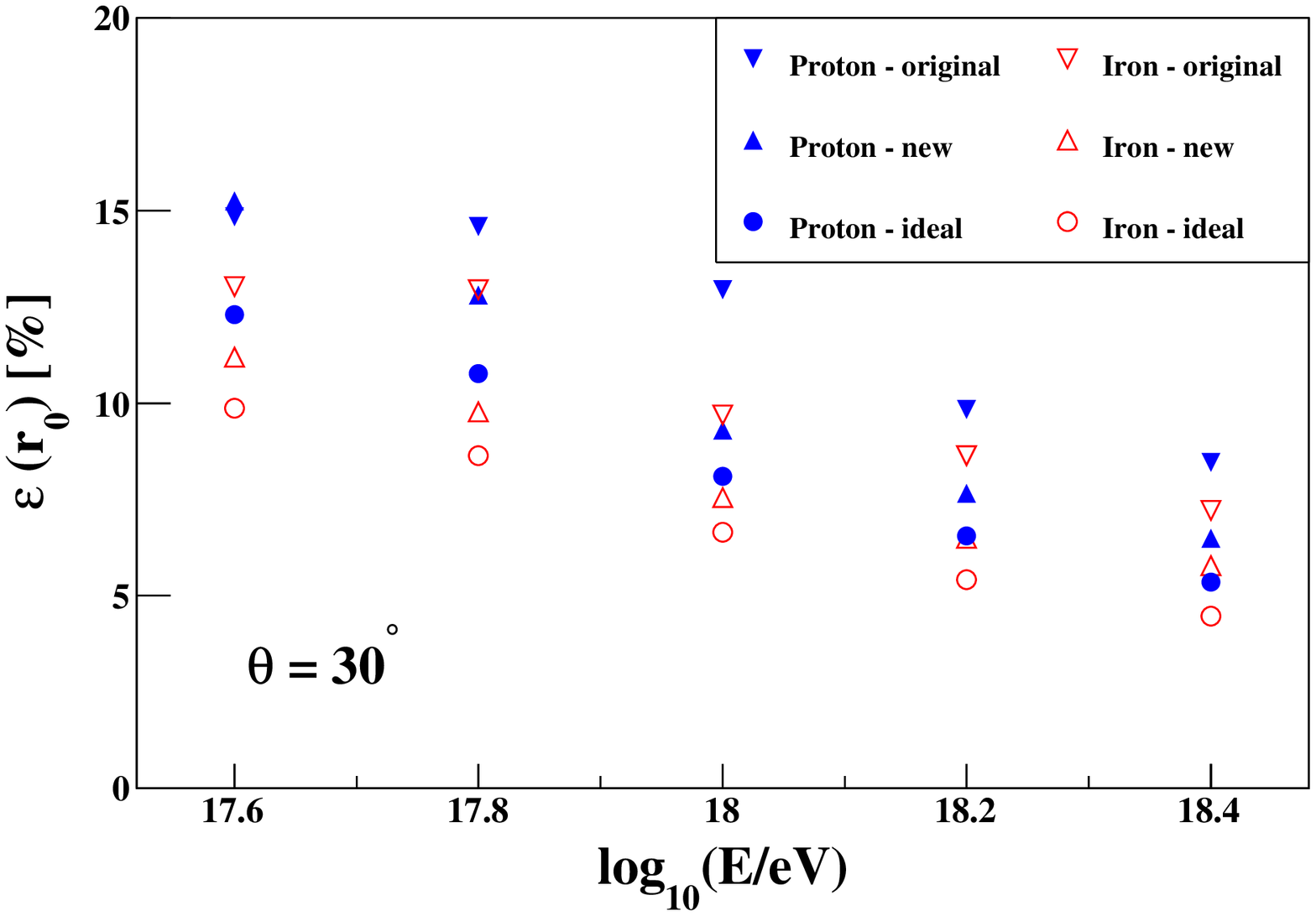}
\includegraphics[width=.45\textwidth]{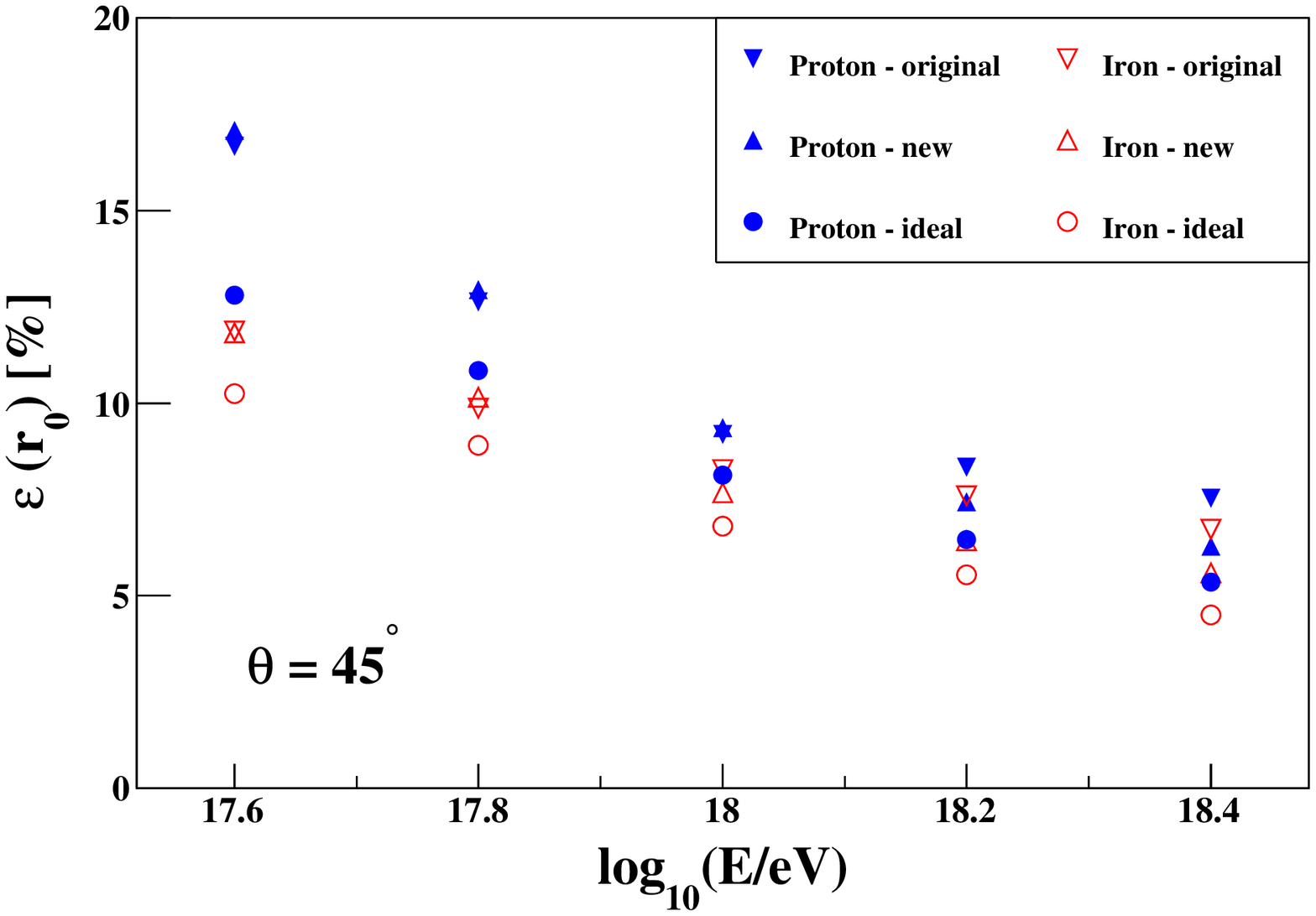}
\caption{Relative standard deviation of the number of muons at $r_0 = \unit[450]{m}$.  Air showers reconstructed with the original reconstruction, the new one, and an ideal muon counter are shown. Zenith angles $30^\circ$ and
$45^\circ$ are displayed in the left and right panels respectively. \label{fig:epsilon}}
\end{figure*}

\section{Performance with energy and zenith angle}

\label{sec:performance}

In this section the performances of the original and the new reconstructions are compared. In general $r_{min}$ depends on primary type, energy, zenith angle, and reconstruction method as shown in Fig.~\ref{fig:rmin}. However a single reference distance ($r_0$) is usually adopted since it is impractical to use a different one for each particular case. We used $r_0 = \unit[450]{m}$ in the original and new reconstructions and in the ideal detector for the reasons explained below. In the new reconstruction and ideal detector $r_0$ is close to the average of the $r_{min}$ of the two primaries and zenith angles at $E=\unit[10^{18}]{eV}$. Although $\varepsilon(r_0)$ is higher than the $\varepsilon(r_{min})$ specific to an energy, zenith angle, and primary particle, the difference is less than 2\% in all simulated air showers. We followed a different criterion for the original reconstruction. We also selected $r_0 = \unit[450]{m}$, but the aim was to minimise $\varepsilon(r_0)$ at low energy. Another important 
consideration is that reconstructions can be compared directly if the same $r_0$ is used in all of them.
The reconstructed MLDF is evaluated at $r_0$ to derive an estimator of the shower size
$\hat{\mu}(r_0)$. The relative bias and standard deviation of $\hat{\mu}(r_0)$ are estimated for each air shower. The bias modulus is less than 3\% and
decreases with energy in the 3 reconstructions as shown in Fig.~\ref{fig:bias}. The $\varepsilon(r_0)$ of the new reconstruction  is lower than in 
the original method in most air showers as shown in Fig.~\ref{fig:epsilon}. However differences in
$\varepsilon(r_0)$ are smaller at $45^\circ$ than at $30^\circ$. 

As mentioned more events are included in the new reconstruction than in the original one given the different quality cuts applied. Therefore, even in air showers with similar $\varepsilon(r_0)$, it is better to use the new method than the original one. 
The $\varepsilon(r_0)$ of the original reconstruction is larger than in the new reconstruction one at $E=\unit[10^{17.6}]{eV}$ and $\unit[10^{17.8}]{eV}$ if the same event sample is used. The increment is $\approx40\%$ for iron at $E=\unit[10^{17.6}]{eV}$ and $\theta=30^\circ$ and even higher for  proton. The rise is caused by outliers in $\hat{\mu}(r_0)$ in the MLDF fit when there is a saturated detector. The problem of a large $\varepsilon(r_0)$ does not happen in the original reconstruction at $E\ge\unit[10^{18}]{eV}$ as shown in Fig.~\ref{fig:epsilon}. For example although 35\% of events saturate the original reconstruction of iron at $E=\unit[10^{18.4}]{eV}$ and $\theta=30^\circ$, $\varepsilon(r_0)$ is only $1.5\%$ larger than in the new reconstruction. The difference is small because, in this shower, non saturated detectors have a sizable signal that stabilises the MLDF fit. The resolution for iron is better than for proton since air showers initiated by the former have more muons. The new method $\
varepsilon(r_0)$ is 
close to the ideal detector lower bound except for protons with energy below $\unit[10^{18}]{eV}$. 

It is already firmly established that there are fewer muons in simulations than in observed air showers~\cite{Farrar:13}.
The deficit is originated by the lack of knowledge of hadronic interactions at the highest energies. In particular, hadronic models used in simulations extrapolate accelerator data to cosmic ray energies. With more muons, $\varepsilon(r_0)$ decreases allowing for a better primary mass discrimination. However, there are also more saturated events. We analysed the effect of more muons by doubling the muon content predicted by QGSJET-II-03 in the average MLDF. Reconstructions of iron at $\theta=30^\circ$ with the new method show that $\varepsilon(r_0)$ is reduced only $\sim3\%$ at $E=\unit[10^{17.6}]{eV}$. Deviations get even smaller with energy, becoming $1\%$ at $\unit[10^{18.4}]{eV}$. Differences in $\varepsilon(r_0)$ are low despite large muon variations because the detector resolution is flat with the signal level as shown in \ref{sec:appendixA}. On the other hand, with twice as many muons unsaturated events are reduced to $76\%$.

\begin{figure}[tp]
\centering
\setlength{\abovecaptionskip}{0pt}
\includegraphics[width=.4\textwidth]{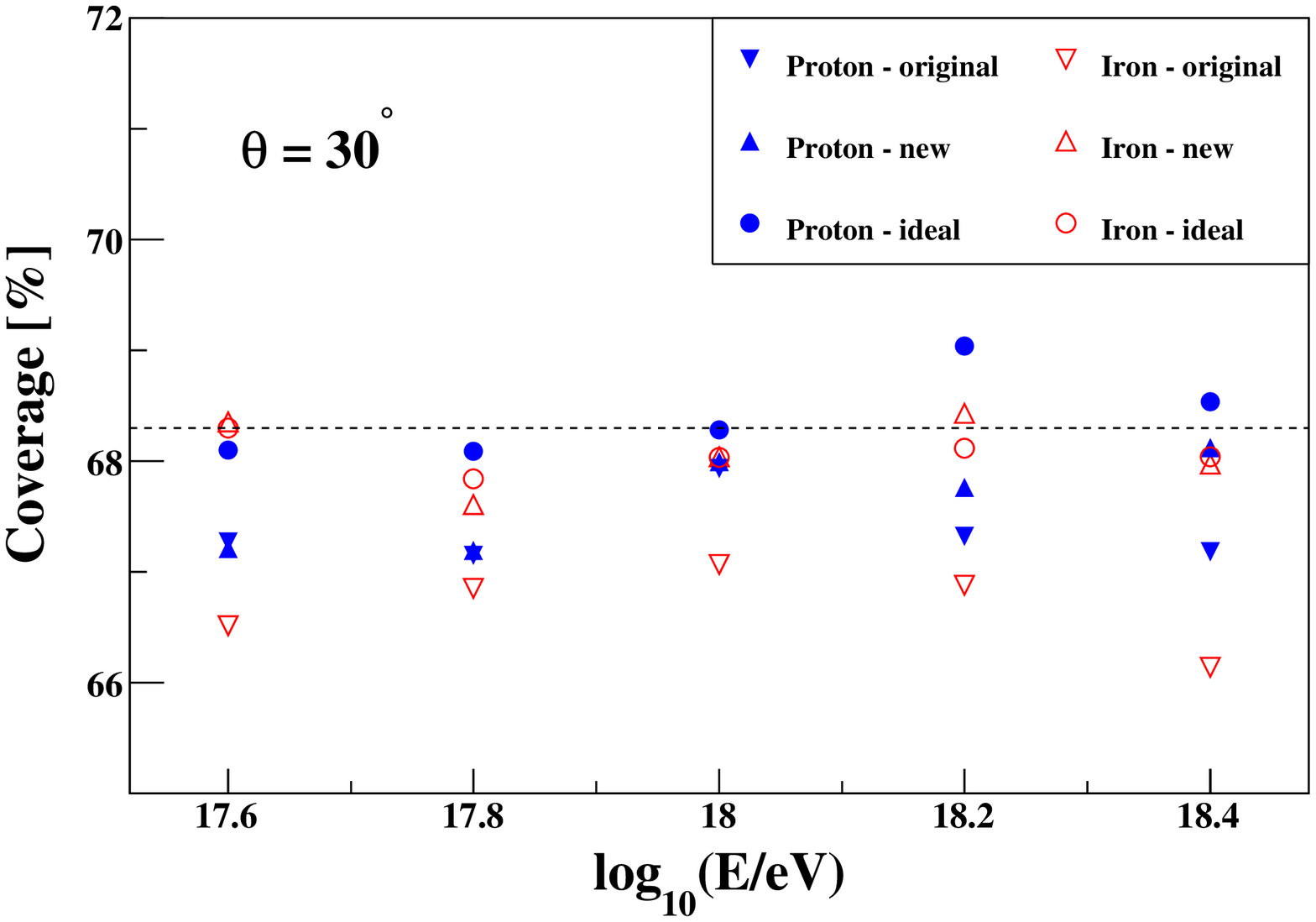}
\includegraphics[width=.4\textwidth]{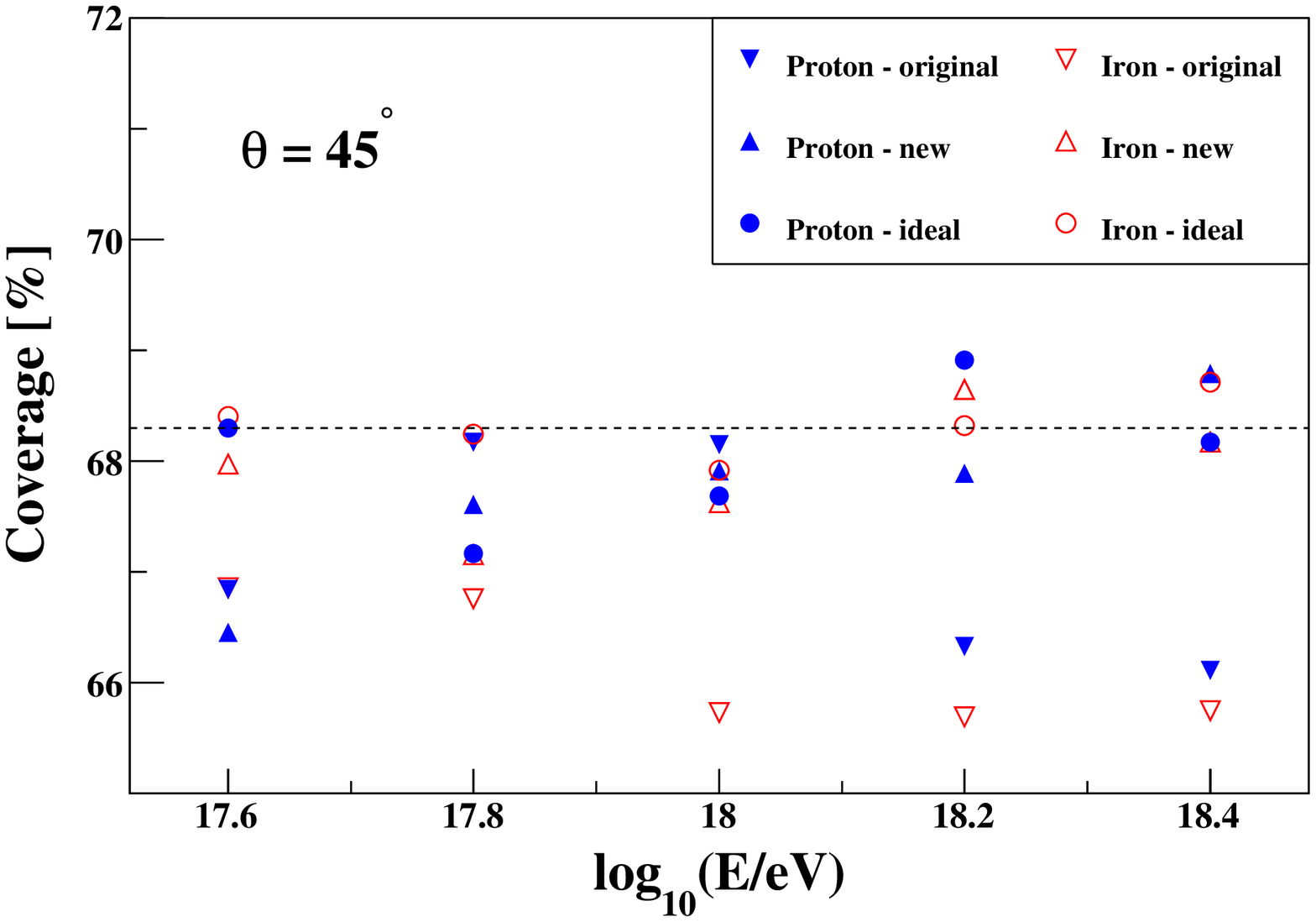}
\caption{Coverage of the 1$\sigma$ confidence interval of the number of muons at \unit[450]{m} for the original reconstruction, the new reconstruction, and an ideal detector. Zenith angles $30^\circ$ and
$45^\circ$ are displayed in the top and bottom panels respectively. The coverage of the
1$\sigma$ Gaussian confidence interval is shown as a dotted line. The statistical error in the coverage is $0.5\%$.
\label{fig:cover}}
\end{figure}

The coverage of the $\mu(r_0)$ confidence interval obtained from the MLDF reconstruction is another useful performance measure. In the reconstruction of real events confidence intervals of fit parameters are calculated from the propagation of data errors. Coverage is defined as the probability that a confidence interval contains the true value of an estimated parameter \cite{Beringer2012}. For example the coverage of the 1$\sigma$ confidence interval of a Gaussian distribution is 68.3\%. In the more general case of a distribution approximately Gaussian the coverage is close to this value. If data errors are underestimated the coverage of a fit parameter can be significantly lower than the Gaussian nominal value. We calculated the coverage of $\mu(r_0)$ as the fraction of events which have a confidence interval that includes the true value taken from the simulated MLDF. The new reconstruction and ideal detector coverages are close to the Gaussian value as shown in Fig.~\ref{fig:cover}. In principle the 
original reconstruction undercovers 
more than the two other methods because data uncertainties are underestimated. In this method good stations include Poissonian fluctuations due to the finite number of muons but not a detector segmentation contribution. The coverages of the original and the new methods are similar below $\unit[10^{18}]{eV}$. At higher energies however, the original reconstruction covers less than the new one. This coverage problem happens because signals are larger at higher energy and, therefore, data errors are underestimated more.

\section{Performance with other configurations}
\label{sec:Conf}

\begin{figure}[htp]
\centering
\setlength{\abovecaptionskip}{0pt}
\includegraphics[width=.4\textwidth]{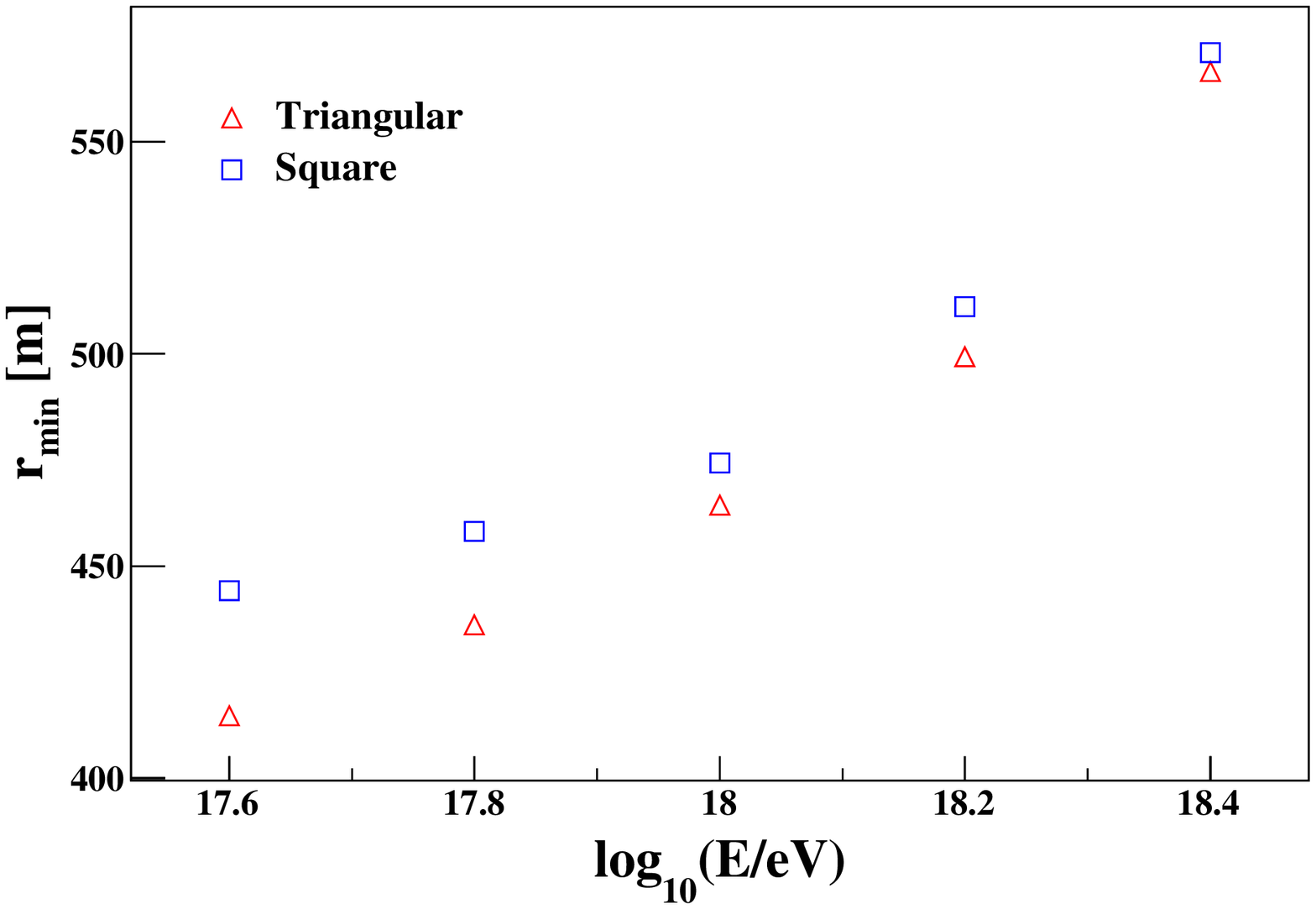}
\includegraphics[width=.4\textwidth]{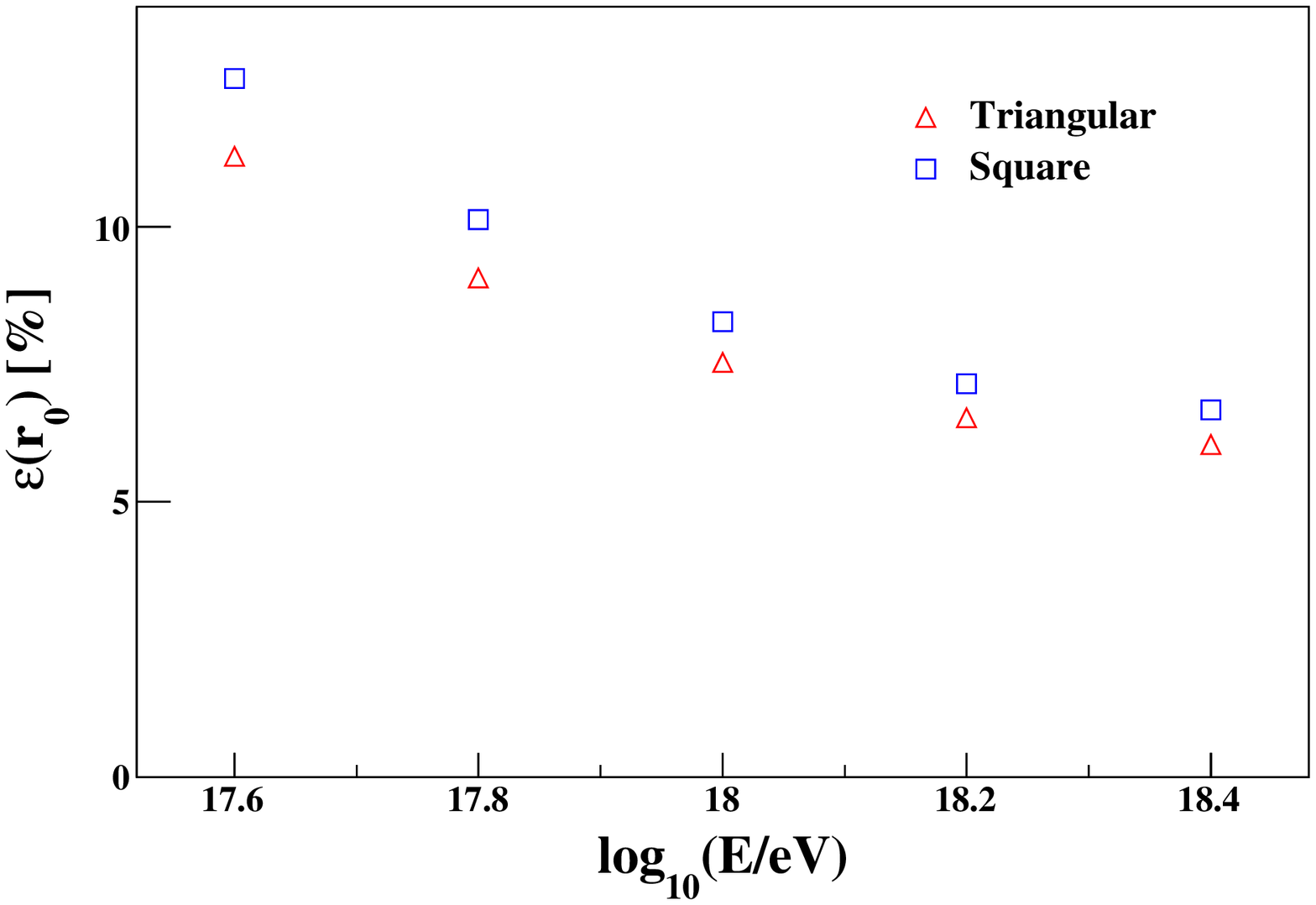}
\caption{Minimal distance (top) and relative standard deviation of the number of muons at $r_0= \unit[450]{m}$
(bottom) as function of energy. The AMIGA triangular array and a square one spaced at \unit[750]{m} are displayed.\label{fig:perfgeo}}
\end{figure}

The performance of the new reconstruction with 3 different array configurations is presented in this section. A square array spaced at \unit[750]{m} is considered first. For this array, iron showers at $\theta=30^{\circ}$ are reconstructed and events are selected using the saturation cut presented in section~\ref{sec:saturation}. The $r_{min}$ of the square array is higher than in the more compact AMIGA triangular array, as shown in the top panel of Fig.~\ref{fig:perfgeo}. At high energy the difference is less because more detectors participate in the reconstruction. For the lowest simulated energies $r_{min}$ is close to \unit[450]{m}. As in the AMIGA triangular array, a single reference distance $r_0= \unit[450]{m}$ is used because $\varepsilon(r_0)$ is close to $\varepsilon(r_{min})$ in all considered energies. The $\varepsilon(r_0)$ of the triangular and the square array is shown in the bottom panel of Fig.~\ref{fig:perfgeo}. The difference between both arrays is less than $1.5\%$. The square array 
bias is negligible and its coverage close to the Gaussian value. 

\begin{figure}[htp]
\centering
\setlength{\abovecaptionskip}{0pt}
\includegraphics[width=.45\textwidth]{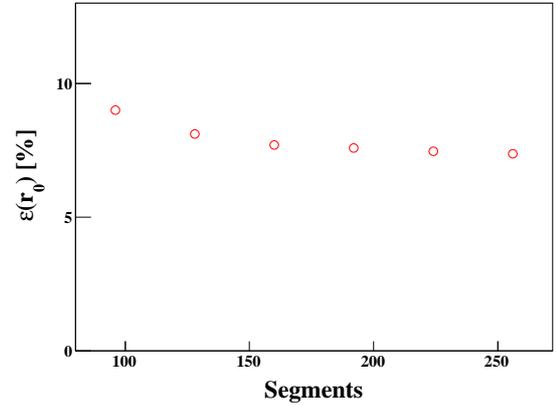}
\caption{Relative standard deviation of the number of muons at the reference distance $r_0=\unit[450]{m}$ as function of the number of
segments. A triangular array is considered.\label{fig:perfseg}}
\end{figure}

The $\varepsilon(r_{0})$ of the new reconstruction varying the number of segments but keeping the same surface area per position, is also assessed. Segmentations from $n=96$ to $n=256$ in steps of $\Delta n = 32$ are compared. Simulations of iron at $E=\unit[10^{18}]{eV}$ and $\theta=30^{\circ}$ with the saturation cut applied are used to compare segmentations. It is remarkable that $\varepsilon(r_{0})$ depends mildly on segmentation, as shown in Fig.~\ref{fig:perfseg}. In this example the resolution is dominated by the finite number of muons rather than by the detector segmentation. The shower has $36.4$ muons in the detector area at $r_0$, fewer than the detector segments in all considered cases. However when there are more segments more events are unsaturated. The fraction of reconstructed events improves from $89\%$ at $n=96$ to $98\%$ at $n=256$.      

\begin{figure}[htp]
\centering
\setlength{\abovecaptionskip}{0pt}
\includegraphics[width=.45\textwidth]{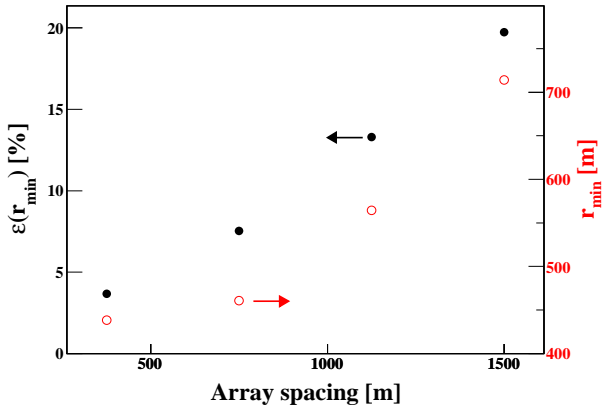}
\caption{Left axis: Relative standard deviation of the number of muons at the minimal distance as function of the array
spacing. Right axis: Minimal distance against the detector spacing.\label{fig:perdist} }
\end{figure}

The third configuration considered is a triangular array with different detectors spacings. Distances between \unit[375]{m} and \unit[1500]{m}, the Auger surface detector spacing, are tested with an iron air shower at $E=\unit[10^{18}]{eV}$ and $\theta=30^{\circ}$. The fraction of events selected with the saturation cut varies from $89\%$ at \unit[375]{m} to $99.6\%$ at \unit[1500]{m}. The \unit[1500]{m} array is close to the reconstruction threshold, $9\%$ of the events are not reconstructed because they have less than 2 detectors with a signal. With the array spacing $r_{min}$ increases as shown in Fig.~\ref{fig:perdist}. Its rise is not linear due to the effect of saturation which is more important when detectors are closer. While $\varepsilon(r_{min})$ grows from $4\%$ at \unit[375]{m} to $21\%$ at \unit[1500]{m}, the bias at $r_{min}$ is less than $1\%$ at all distances. 

\section{Conclusions}

\label{sec:conclusion}

In this work we presented a new method for reconstructing the muon lateral distribution function with an array of segmented counters. The new reconstruction is based on the exact likelihood of the number of muons in a detector given the number of segments with a signal. The combined saturation limit of the 3 counters at each array position increases from 174 muons per \unit[25]{ns} time bin in the original reconstruction to 719 muons for the whole event duration in the new reconstruction. The high muon signals causing saturation are also short because they are close to the air shower core. In this case most muons arrive in a single time bin. The original reconstruction saturation, valid for each time bin, can be then compared directly to the new reconstruction. The rise of the saturation threshold in the new method allows for the reconstruction of air showers falling closer to a detector.  As a result the number of events that can be used for science analyses is increased.

We found an optimal distance of \unit[450]{m} to estimate the number of muons. At this distance statistical fluctuations of the fitted lateral distribution function are minimised. The found reference distance coincides with the \unit[750]{m} surface detector array of the Pierre Auger Observatory~\cite{Ravignani2013}. The statistical uncertainty of the new reconstruction is lower than in the original method. This resolution improvement will allow for a better cosmic ray mass classification.

The original and new reconstruction systematic biases can be neglected since they are much lower than the  corresponding statistical uncertainties. The confidence interval coverage of the number of muons at \unit[450]{m} is close to the value expected from a Gaussian distribution in the new reconstruction. In the original method there is an undercoverage of $\sim2\%$ consistent with the underestimation of data errors. The resolution achieved with the new method is only $\sim1\%$ higher, in average, than the limit set by an ideal muon counter. Therefore there is little room for improvement to make by adding the detector time resolution. 

We also assessed the new reconstruction performance with 3 different array configurations by comparing statistical uncertainties in the number of muons. We showed that they are similar for the AMIGA triangular array and a square one spaced at \unit[750]{m}. We also found that the statistical uncertainty is weakly dependent on the segmentation for an air shower of iron at $E=\unit[10^{18}]{eV}$ and $\theta=30^\circ$. Finally we showed that the statistical uncertainty depends on the array spacing and ranges from 4\% at \unit[375]{m} to 21\% at \unit[1500]{m}.

\appendix

\section{Uncertainty of the muon density measured with a segmented detector}

\label{sec:appendixA}

\begin{figure}[htp]
\centering
\setlength{\abovecaptionskip}{0pt}
\includegraphics[width=.45\textwidth]{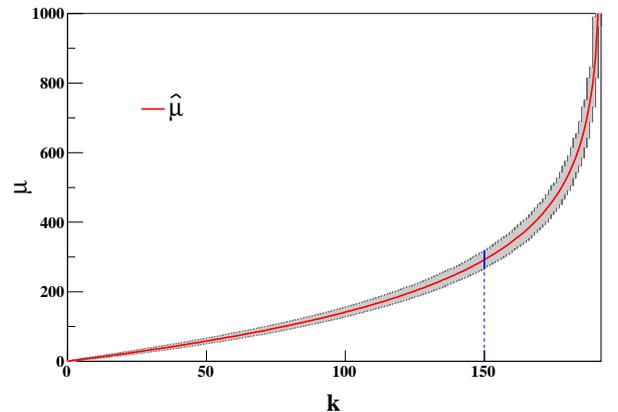}
\caption{ Maximum likelihood estimator of the number of expected muons $\hat{\mu}$ as function of the number of segments
with a signal $k$ (red line) and associated 1$\sigma$ Feldman-Cousins confidence belt (grey area). The confidence
interval for $k=150$ is shown with a solid blue line. \label{fig:cbelt}}
\end{figure} 

The 1$\sigma$ confidence interval of the density measured with a segmented counter is presented in this appendix. The
example of a detector divided into 192 segments like the AMIGA muon counters is used. However the outlined method applies to
any number of segments and any kind of particles. The number of expected particles $\mu$ is proportional to the density
(see Eq.~(\ref{eq:mu})). The estimator of maximum likelihood $\hat\mu$ was already introduced in Eq.~(\ref{eq:muest}).
This estimator is finite when $k$ is less than the number of segments and it diverges when all the segments have a signal
(i.e., $k$ = n ). $\hat{\mu}$ as a function of the number of segments with a signal is shown in Fig.~\ref{fig:cbelt}. The
1$\sigma$ confidence belt shown in this figure is obtained using the Neyman construction in steps $\Delta\mu=0.15$ from 
$\mu = 0$ to $1500$. For each $\mu$ an acceptance interval limited by two values of $k$, $k_{min}$ and $k_{max}$, is 
defined as
\begin{equation}
P(k_{min} \le k \le  k_{max}; \mu ) \ge 1 - \alpha = 0.683.
\label{eq:cli}
\end{equation}   

$1-\alpha$ is the probability contained in a 1$\sigma$ interval of a Gaussian distribution. A
minimum for the probability instead of a strict equality is used in Eq.~(\ref{eq:cli}) due to the discretisation of $k$.
The procedure proposed by Feldman and Cousins in \cite{Feldman1998} is used to find the confidence intervals. This
method is favoured over other possible options, as the standard central interval, because the overcoverage is reduced to a level consistent with the discretisation of $k$. When only 1 segment has a signal, $\hat{\mu}=1$ and the 1$\sigma$ confidence interval is $[0.6,1.7]$.

\begin{figure}[htp]
\centering
\setlength{\abovecaptionskip}{0pt}
\includegraphics[width=.45\textwidth]{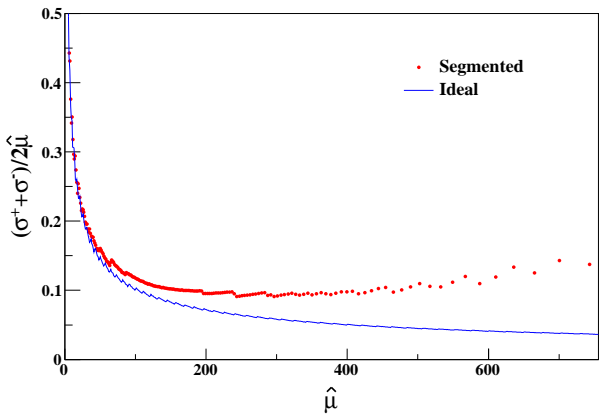}
\caption{Resolutions of a segmented detector and an ideal counter as a function of the estimated number of muons $\hat{\mu}$. In this example the detector has 192 segments.
\label{fig:reso}}
\end{figure}

The counter resolution is driven by its segmentation and by the number of collected muons given its size. It is calculated here as the half-length of the confidence interval over $\hat{\mu}$. The
resolutions of the segmented detector and of an ideal counter are shown in Fig.~\ref{fig:reso}. The resolution of the
segmented detector is notably flat at $\approx 10\%$ in a wide range of $\hat{\mu}$. When there are few segments on the
resolution becomes poorer due to the low number of muons in the detector. The resolution also deteriorates close to
saturation as an effect of the detector segmentation. 

The ideal muon counter resolution is caused by the Poissonian fluctuations of the number of muons. The
segmented detector has a resolution close to the ideal counter if $\mu$ is much less than the number of segments. When
$\mu$ becomes larger the segmentation causes the resolution to be lower than in the ideal detector. For example the 1$\sigma$ interval of a segmented counter with 150 segments on is $292 \pm 27$, whereas the interval for an ideal detector is $292 \pm 18$. 

\section*{Acknowledgements}
This work was informed by very helpful discussions with A. Etchegoyen. The authors have also greatly benefited from
discussions with several colleagues from the Pierre Auger Collaboration, of which they are members. We specially thank
St\'ephane Coutu and Carola Dobrigkeit for the review of the manuscript. A. D. S. is a member of the Carrera del Investigador
Cient\'{\i}fico of CONICET, Argentina. This work was partially funded by PIP 114-201101-00360 (CONICET) and PICT
2011-2223 (ANPCyT).

\end{document}